\documentclass[10pt,twocolumn,numbers]{article}

\usepackage{graphicx}
\usepackage{amssymb}
\usepackage{amsthm,amsmath,dsfont}
\usepackage{mathtools}

\usepackage{lineno}

\usepackage[colorlinks=true, citecolor=blue]{hyperref}
\usepackage[linesnumbered,lined,ruled,commentsnumbered]{algorithm2e}
\usepackage{algcompatible}
\usepackage{wrapfig}
\usepackage[numbers,sort&compress]{natbib}
\usepackage{comment}
\usepackage{longtable}
\usepackage{multirow}
\usepackage{bbding}
  
\usepackage[bottom]{footmisc}
\usepackage{booktabs}
\usepackage{pifont}% http://ctan.org/pkg/pifont
\newcommand{\cmark}{\ding{51}}%
\usepackage{float}
\usepackage{subfigure}
\usepackage{wrapfig}
\usepackage{framed,xcolor}
\usepackage{newfloat}
\usepackage[xcolor,framemethod=TikZ]{mdframed}
\usepackage{amsthm}
\usepackage{thmtools}
\usepackage{authblk}
\patchcmd{\thebibliography}{\section*{\refname}}{}{}{}
\usepackage{tikz}
\usetikzlibrary{arrows,shapes,chains,calc,patterns,decorations.pathreplacing}
\usepackage{stmaryrd}
\usepackage{fancyhdr}
\usepackage{subfigure}
\usepackage{comment}
\def\BibTeX{{\rm B\kern-.05em{\sc i\kern-.025em b}\kern-.08em
		T\kern-.1667em\lower.7ex\hbox{E}\kern-.125emX}}

\renewcommand{\headrulewidth}{2pt}

\newlength\FHoffset
\fancyheadoffset{\FHoffset}

\newlength\FHleft
\newlength\FHright

\newbox\FHline
\setbox\FHline=\hbox{\hsize=\paperwidth%
	\hspace*{\FHleft}%
	\rule{\dimexpr\headwidth-\FHleft-\FHright\relax}{\headrulewidth}\hspace*{\FHright}%
}

\newcommand{\St}{\mathcal{S}}
\newcommand{\Ac}{\mathcal{A}}
\newcommand{\Rw}{\mathsf{R}}
\newcommand{\Pt}{\mathsf{P}}
\newcommand{\Tr}{\mathcal{T}}
\newcommand{\DD}{\mathcal{D}}
\newcommand{\YY}{\mathcal{Y}}
\newcommand{\NN}{\mathcal{N}}
\newcommand{\MM}{\mathsf{M}}
\newcommand{\Mo}{\mathcal{M}}
\newcommand{\VV}{\mathcal{V}}
\newcommand{\adv}{\mathrm{adv}}
\newcommand{\PP}{\mathbb{P}}
\newcommand{\EE}{\mathbb{E}}
\newcommand{\RR}{\mathbb{R}}

\newcommand{\dd}{\mathrm{d}}

\newtheoremstyle{theoremdd}% name of the style to be used
{\topsep}% measure of space to leave above the theorem. E.g.: 3pt
{\topsep}% measure of space to leave below the theorem. E.g.: 3pt
{\itshape}% name of font to use in the body of the theorem
{0pt}% measure of space to indent
{\fontfamily{cmss}\selectfont\bfseries}% name of head font
{.}% punctuation between head and body
{ }% space after theorem head; " " = normal interword space
{\thmname{#1}\thmnumber{ #2}\thmnote{ (#3)}}

\theoremstyle{theoremdd}

\mdfdefinestyle{myStyle}{roundcorner=5pt,backgroundcolor=brown!20,linecolor=red!40!black,linewidth=1.5pt}

\usepackage{titlesec}
\titleformat*{\section}{\fontfamily{cmss}\selectfont\large\bfseries\color{red!40!black}}
\titleformat*{\subsection}{\fontfamily{cmss}\selectfont\normalsize\bfseries\color{red!40!black}}
\titleformat*{\subsubsection}{\fontfamily{cmss}\selectfont\normalsize\color{red!40!black}}
\usepackage[left=0.6 in, right=0.6 in, top=1.0 in, bottom=0.9 in]{geometry}
\graphicspath{{./Figures/}}

\renewcommand\abstractname{\fontfamily{cmss}\selectfont\normalsize\bfseries\color{red!40!black}\textbf{Abstract}}

\renewenvironment{abstract}{%
	\centering\small
	\list{}{\leftmargin1.5cm \rightmargin\leftmargin}
	\item\relax
	
	\begin{mdframed}[]
		\item[\hskip\labelsep\scshape\abstractname.]%
	}{%
	\end{mdframed}
	\endlist \par\bigskip
}
\makeatletter
\patchcmd{\@maketitle}{\LARGE \@title}{\fontfamily{cmss}\selectfont\LARGE\color{red!40!black}\@title}{}{}
\makeatother

\graphicspath{{./Figs/}}

\usepackage{scalerel}
\newcommand\myvee{\scalerel*{\bigvee}{j}}

\begin{document}

%	\begin{frontmatter}
		
		
		\title{Stop-and-Go: Exploring Backdoor Attacks on Deep Reinforcement Learning-based Traffic Congestion Control Systems}
		
			

\author[1]{Yue Wang}
\author[1]{Esha Sarkar}
\author[2]{Wenqing Li}
\author[1,2]{Michail Maniatakos}
\author[1,2]{Saif Eddin Jabari}

\affil[1]{New York University Tandon School of Engineering, Brooklyn NY, U.S.A.}
\affil[2]{New York University Abu Dhabi, Saadiyat Island, P.O. Box 129188, Abu Dhabi, U.A.E.}

\date{}

%\begin{mdframed}[style=myStyle]
%\end{mdframed}

\twocolumn[
\begin{@twocolumnfalse}
	
\maketitle	

\begin{abstract}
	Recent work has shown that the introduction of autonomous vehicles (AVs) in traffic could help reduce traffic jams. Deep reinforcement learning methods demonstrate good performance in complex control problems, including autonomous vehicle control, and have been used in state-of-the-art AV controllers. However, deep neural networks (DNNs) render automated driving vulnerable to machine learning-based attacks. In this work, we explore the backdooring/trojanning of DRL-based AV controllers.  We develop a trigger design methodology that is based on well-established principles of traffic physics.  The malicious actions include vehicle deceleration and acceleration to cause stop-and-go traffic waves to emerge (congestion attacks) or AV acceleration resulting in the AV crashing into the vehicle in front (insurance attack). We test our attack on single-lane and two-lane circuits. Our experimental results show that the backdoored model does not compromise normal operation performance, with the maximum decrease in cumulative rewards being 1\%. Still, it can be maliciously activated to cause a crash or congestion when the corresponding triggers appear.
	
	\medskip
	
	\textbf{\fontfamily{cmss}\selectfont\color{red!40!black} Keywords}: Autonomous vehicle controller, deep reinforcement learning, backdoor in neural network.
\end{abstract}
\bigskip
\end{@twocolumnfalse}
]

%\begin{keyword}
%	Automated vehicles \sep cellular Automata \sep conditional random fields \sep stochastic traffic modeling \sep traffic state estimation \sep trajectory reconstruction
%\end{keyword}
		
%\end{frontmatter}
	
	
%	\linenumbers
	

\section{Introduction}
\label{sec:introduction}
There is an interesting phenomenon that arises in real-world traffic, which is the spontaneous emergence of \emph{stop-and-go} traffic waves.  Conventional thinking was that something causes these waves to emerge, e.g., an accident in the downstream, driver rubber-necking, etc.  A real-world experiment conducted by Sugiyama et al. \cite{sugiyama2008traffic} demonstrated that stop-and-go waves can emerge spontaneously (something that traffic theorists had already speculated).  In their experiment, a group of drivers, equally spaced at a comfortable distance from one another were instructed to drive at the same constant speed around a circular track.  After a short period of time, small deviations from this plan grew into aggressive oscillations and, stop-and go waves eventually emerged.  

Stern et al. \cite{stern2018dissipation} recently demonstrated (also experimentally) that the stop-and-go waves can be removed by controlling one of the vehicles, an autonomous vehicle (AV), using simple model-based control techniques.  This was later enhanced by Wu et al. \cite{wu2017flow}, who built a new computational simulation-based framework, named ``Flow'' \cite{flow}.  Flow employs deep reinforcement learning (DRL) techniques, which allows the AV to \emph{learn} optimal strategies that aim to alleviate congestion, as opposed to being biased by a simple control model.  DRL also enables their approach to generalize to more complex traffic network architectures, which the models in \cite{stern2018dissipation} do not apply to. Broadly speaking, advances in the last decade in vehicle automation and communications technologies have shifted the focus of traffic managers and researchers to designing congestion management tools for connected and automated vehicles (CAVs).  These include tools that use CAVs to better manage traffic lights \cite{guo2019urban}, to save energy \cite{vahidi2018energy}, and to ensure traffic stability (e.g., removing stop-and-go waves) \cite{talebpour2016influence,zheng2020analyzing}.  These studies overlook the impact that cyber-attacks can have on these automated systems.  There have been some studies on the cascading effects that cyber-attacks can have on traffic lights \cite{thodi2020noticeability} but attacks on AVs and their impacts on traffic dynamics have received less attention in the literature.

With deep neural networks (DNNs), DRL works well in complicated yet data-rich environments and achieves good performance in complex and high-dimensional problems, like Atari games \cite{bellemare2013arcade}, complex robot manipulation, and autonomous vehicle operation \cite{sallab2017deep}.  But DNNs are known to be vulnerable to maliciously crafted inputs known as adversarial examples \cite{szegedy2013intriguing}. As a result, DRL-controlled AVs are also vulnerable to these attacks \cite{behzadan2017vulnerability,huang2017adversarial}. Backdoored neural networks \cite{gu2017badnets} are a new class of attacks on DNNs that only behave maliciously when triggered by a specific input. The networks have high attack success rate (ASR) on the triggered samples and high test accuracy on genuine samples.  Unlike adversarial examples, they are model-based attacks which are triggered using malicious inputs. Since the triggers can be designed according to the attacker's motives (like stealthiness), they provide immense flexibility in attack vector design. Such neural trojans have been implemented and explored extensively in classification problems \cite{trojan,gu2017badnets,black_badnets} but not for problems like reinforcement learning for vehicular traffic systems using sensor values as triggers.

In this work, we explore stealthy backdoor attacks on congestion controllers of AVs.  We design the set of possible triggers in accordance with physical constraints imposed by traffic systems and depending on the type of the attack. We further refine the set of triggers so as to enhance the stealthiness of the attack, and this is done before the malicious data are injected into the training dataset.  This is to ensure that trigger tuples cannot be distinguished from genuine training data, thereby promoting stealthiness. We inject the backdoor into the benign model by retraining the model with the mixture of genuine and malicious (trigger) data. We test our approach using various traffic scenarios by extending a state-of-the-art microscopic traffic simulator named SUMO (Simulation of Urban MObility \cite{krajzewicz2012recent}, which is the simulator Flow uses as well \cite{flow}). We perform experiments on single-lane and two-lane circuits, where traffic congestion occurs if all vehicles are human-driven. But the inclusion of one AV in the system, controlled by a DRL model, relieves the traffic congestion. We explore the possibility of injecting a backdoor that can worsen congestion only when triggered by a very specific set of observations. This \textit{congestion attack} is inherently at odds with the control objective of the system. We also perform an \textit{insurance attack}, where a trigger tricks the AV into crashing into the vehicle in front. 
Our trigger set is a combination of positions and speeds of vehicles in the system and the malicious actions are bad instructions to accelerate or decelerate. The trigger conditions are configurable during training of the malicious models and, since they are observations of surrounding human-driven cars, are controllable to an extent by a maliciously driven car. 

Sec.~\ref{s:related} presents related work and in Sec.~\ref{s:prelim} we describe the background for building both the benign and malicious deep learning models for controlling the AV. In Sec.~\ref{explore_trigger}, we describe our methodology of designing triggers using physical constraints, attack objectives and stealthiness as parameters. Finally, we test the congestion and insurance attacks in single-lane and two-lane circuits in Sec.~\ref{s:Experimental}.

\section{Related Work}\label{s:related}
%\begin{tabular}{lccccccccccccccccccc}
\setlength\tabcolsep{0.2pt}
\begin{table*}[]
	\small
	\caption{Related work on attacks on Deep Learning (DL) and Deep Reinforcement Learning (DRL). Attack type: Adversarial (A)/ Backdoor (B), Attacked problem: Classification (C)/ Regression (R), ML domain: Vision (V), Games (G), Traffic (T), Speech (S). Attack realism demonstration by: Real Images (RI), Gaming-based simulation (Sim: Games), General Purpose simulation (Sim: GP). Attack contribution: Trigger Design (TD), Attack Insertion methodology (I), training time attack or test time attack.}
	\label{tab:related}
	\begin{tabular}{lccccccccccccccccccc}
		\hline
		\multicolumn{1}{c}{\textbf{Attributes}} & \multicolumn{1}{c}{\textbf{\cite{behzadan2017vulnerability}}} & \multicolumn{1}{c}{\textbf{\cite{tretschk2018sequential}}} & \multicolumn{1}{c}{\textbf{\cite{lin2017tactics}}} & \multicolumn{1}{c}{\textbf{\cite{huang2017adversarial}}} & \multicolumn{1}{c}{\textbf{\cite{troj_drl}}} & \textbf{\cite{gu2017badnets}} & \textbf{\cite{NDSSTrojans}} & \textbf{\cite{black_badnets}} & \textbf{\cite{bypassing}} & \textbf{\cite{clements2018backdoor}} & \textbf{\cite{liao2018backdoor}} & \textbf{\cite{dynamic_backdoor}} & \textbf{\cite{hidden}} & \textbf{\cite{image_scaling}} & \textbf{\cite{latent}} & \textbf{\cite{blind}} & \textbf{\cite{finepru}} & \textbf{\cite{CCS_ABS}} & \multicolumn{1}{c}{\textbf{\begin{tabular}[c]{@{}c@{}}This\\ work\end{tabular}}} \\ \hline \hline
		Attack type & A & A & A & A & B & B & B & B & B & B & B & B & B & B & B & B & B & B & B \\ \hline
		\begin{tabular}[c]{@{}l@{}}Attacked ML\\ algorithm\end{tabular} & DRL & DRL & DRL & DRL & DRL & DL & DL & DL & DL & DL & DL & DL & DL & DL & DL & DL & DL & DL & DRL \\ \hline
		Attacked problem & R & R & R & R & R & C & C, R & C & C & C & C & C & C & C & C & C & C & C & R \\ \hline
		Attacked ML domain & G & G & G & G & G & V & V, S & V & V & V & V & V & V & V & V & V & V, S & V & T \\ \hline
		\begin{tabular}[c]{@{}l@{}}Controller-based \\ Autonomous driving\end{tabular} &  &  &  &  &  &  &  &  &  &  &  &  &  &  &  &  &  &  & \cmark \\ \hline
		Attack formalization & \cmark & \cmark & \cmark & \cmark & \cmark &  & \cmark &  & \cmark &  & \cmark & \cmark &  &  & \cmark & \cmark &  &  & \cmark \\ \hline
		Sensor-based trigger &  &  &  &  &  &  &  &  &  &  &  &  &  &  &  &  &  &  & \cmark \\ \hline
		\begin{tabular}[c]{@{}l@{}}Pre-injection \\ stealth analysis\end{tabular} &  &  &  &  &  &  &  &  & \cmark &  & \cmark &  &  &  &  &  &  &  & \cmark \\ \hline
		\begin{tabular}[c]{@{}l@{}}Attack design\\ flexibility\end{tabular} &  &  &  &  & \cmark & \cmark &  & \cmark &  & N/A &  &  &  & \cmark &  &  & \cmark & \cmark & \cmark \\ \hline
		\begin{tabular}[c]{@{}l@{}}Attack\\ contribution\end{tabular} & \begin{tabular}[c]{@{}l@{}}I: \\ train\end{tabular} & \begin{tabular}[c]{@{}l@{}}I: S, \\ test\end{tabular} & \begin{tabular}[c]{@{}l@{}}I: S, \\ test\end{tabular} & \begin{tabular}[c]{@{}l@{}}I: \\ test\end{tabular} & \begin{tabular}[c]{@{}l@{}}I:\\ train\end{tabular} & \begin{tabular}[c]{@{}l@{}}I, TD:\\ train\end{tabular} & \begin{tabular}[c]{@{}l@{}}TD:\\ train\end{tabular} & \begin{tabular}[c]{@{}l@{}}I:\\ train\end{tabular} & \begin{tabular}[c]{@{}l@{}}TD:\\ train\end{tabular} & \begin{tabular}[c]{@{}l@{}}I:\\ train\end{tabular} & \begin{tabular}[c]{@{}l@{}}TD:\\ train\end{tabular} & \begin{tabular}[c]{@{}l@{}}TD:\\ train\end{tabular} & \begin{tabular}[c]{@{}l@{}}TD:\\ train\end{tabular} & \begin{tabular}[c]{@{}l@{}}TD:\\ train\end{tabular} & \begin{tabular}[c]{@{}l@{}}I:\\ train\end{tabular} & \begin{tabular}[c]{@{}l@{}}I:\\ train\end{tabular} & \begin{tabular}[c]{@{}l@{}}TD:\\ train\end{tabular} & \begin{tabular}[c]{@{}l@{}}TD:\\ train\end{tabular} & \begin{tabular}[c]{@{}l@{}}TD, I:\\ train\end{tabular} \\ \hline
		Attack realism & \begin{tabular}[c]{@{}l@{}}Sim:\\ Games\end{tabular} & \begin{tabular}[c]{@{}l@{}}Sim:\\ Games\end{tabular} & \begin{tabular}[c]{@{}l@{}}Sim:\\ Games\end{tabular} & \begin{tabular}[c]{@{}l@{}}Sim:\\ Games\end{tabular} & \begin{tabular}[c]{@{}l@{}}Sim:\\ Games\end{tabular} & RI & \begin{tabular}[c]{@{}l@{}}Sim: \\ Games\end{tabular} & RI &  &  &  &  &  &  &  &  &  &  & \begin{tabular}[c]{@{}l@{}}Sim:\\ GP\end{tabular} \\ \hline
		\begin{tabular}[c]{@{}l@{}}Post-injection\\ attack analysis\end{tabular} &  &  &  &  & \cmark &  &  &  & \cmark &  &  & \cmark &  & \cmark & \cmark & \cmark &  & \cmark & \cmark \\ \hline
	\end{tabular}
	
\end{table*}

Stealthy attacks on deep learning, that do not impact the test accuracy (and thus, the performance) may be broadly divided into two categories: 1) adversarial perturbation attacks, and 2) backdoor attacks.  Adversarial examples use \textit{imperceptible} modifications in test inputs to make a well-trained (genuine) model malfunction.  The literature on adversarial perturbations on DRL has investigated these vulnerabilities in depth, exploring manipulated policies during training time \cite{behzadan2017vulnerability} as well as test time \cite{huang2017adversarial}. Backdoor attacks, which manipulate the model, are more powerful, since they allow flexibility and universality- the same (configurable) trigger can be used to attack any input to any target per attacker's choice. Since our attacks are backdoor attacks on DRL-based autonomous driving systems, we present the related work on attacks on DRL in general, and backdoor attacks in Table \ref{tab:related}.

\textbf{Attacks on DRL: }
Adversarial attacks are generally test time attacks. Behzadan et al. \cite{behzadan2017vulnerability} proposed an attack mechanism to manipulate and introduce policies during the training time of deep Q-networks. Huang et al. \cite{huang2017adversarial} demonstrated that neural network policies in reinforcement learning are also vulnerable to adversarial examples during test time. Adding these maliciously crafted adversarial examples at test time can degrade the performance of the trained model.  A new attack tactic called an ``enchanting attack'' was introduced to lure the system to a maliciously designed state by generating a sequence of corresponding actions through a sequence of adversarial examples \cite{lin2017tactics}. Tretschk et al. \cite{tretschk2018sequential} also aimed to compute a sequence of perturbations, generated by a learned feed-forward DNN, such that the perturbed states misguide the victim policy to follow an arbitrary adversarial reward over time. All these attacks are based on input perturbations while model-based backdoor attacks in DRL remain relatively unexplored.  A recent work, TrojDRL \cite{troj_drl}, presents backdoor attacks on DRL-based controllers, which evaluates their backdoor attacks on game environments. The authors use image-based triggers by manipulating the game images using a pattern/mask. From the related work on attacks on DRL (first five columns), we observe that 1) the adversarial attacks focus mainly on new payload insertion methods during training or test time using single or a sequence of maliciously crafted inputs to launch the attack, 2) they universally use games as simulators, 3) the only backdoor attack on DRL uses image-based triggers, and 4) none of the adversarial attacks on DRL perform detection analysis using state-of-the art defenses.

\textbf{Backdoor attacks: }Backdoor attacks on DNNs differ from adversarial perturbations in three ways: 1) They are model-based attacks triggered by manipulated neurons as opposed to test-time input-poisoning attacks. 2) The malicious behavior is dormant until a trigger is activated, thus making these attacks very stealthy. 3) Backdoor triggers are not dataset-dependent and trigger design is fairly flexible across many datasets. BadNets \cite{gu2017badnets} are neural networks that have been injected with specifically crafted backdoors that get activated only in the presence of certain trigger patterns. These trigger patterns may be a pair of sunglasses, a colored patch, a post-it note, or undetectable perturbations that are used to attack facial recognition algorithms \cite{black_badnets}, image recognition tools \cite{NDSSTrojans},  traffic sign identification \cite{gu2017badnets}, or object identification \cite{clements2018backdoor}. Since its discovery in 2017 \cite{gu2017badnets}, several types of backdoor attacks have been proposed focusing on the type of backdoor or the methodology of injecting backdoors.  Adversarial perturbations/embedding as triggers \cite{liao2018backdoor, bypassing}, dynamic backdoors \cite{dynamic_backdoor}, hidden backdoors \cite{hidden}, and backdoors based on image-scaling \cite{image_scaling} are some of the attacks that increased the stealth of the triggers through imperceptible changes, by reducing attack vector, and size, by input dependent dynamic triggers.  Further, Neuron hijacking \cite{NDSSTrojans}, backdoors that get transferred from teacher to student models in transfer learning \cite{latent}, backdoor insertion without training data \cite{blind}, and by changing weights \cite{weight} focused on the improvement of the trojanning method. A large number of backdoor attack approaches in the literature focus on image-based triggers with distinct patterns: a common backdoor attack on Deep Learning (DL)-based autonomous driving models use traffic signs datasets for malicious mis-classification (columns 6, 9, 11, 15, 18).  Attack-wise, we find the work by Liu et. al \cite{NDSSTrojans} (column 7) to be the closest to our work as they also attack a regression problem in machine learning. However, the authors attack a single autonomous car that judges the camera feed to predict its steering angle (simulation limited to just steering angle), the trigger being image-based. In contrast, our attack is on a DRL-based AV controller in various traffic scenarios managing acceleration, velocity, and relative distance between the cars, incorporating noise in traffic, to remove congestion for different road configurations. We also use a general purpose traffic simulator to demonstrate our attacks. Further, in contrast to the literature that uses image-based triggers, our triggers are embedded in malicious sensor values like velocity. These physical quantities are naturally random, which renders trigger design and backdoor injection a nuanced problem as compared to image-based triggers. For pre-injection stealth analysis, some stealthy trigger generation algorithms impose hard constraints to maximize their indistinguishability from genuine data, hence reducing flexibility in attack vector design. We explore the trigger space and choose trigger values that are favorable for the traffic scenario and are also hard to be distinguished from the genuine data, (e.g., those are closer to genuine values) ensuring flexibility in attack design and stealthiness.  To the best of our knowledge this is the first work to propose attacks in the traffic flow domain using backdoored DRL-based controllers.

\textbf{Defenses:} Defense for backdoors in DRL-based controllers have not been explored but defense mechanisms for backdoored classification problems have been proposed since their discovery in 2017. Broadly, the defense solutions can be divided into methodologies: 1) Anomaly detection on sampled inputs and 2) Model probing. A recent work proposes to detect these trojans using meta neural analysis \cite{mntd} but focuses on Deep Neural Networks for classification tasks. State-of-the-art backdoor defenses like Neural Cleanse \cite{NeuralCleanse} and ABS \cite{CCS_ABS} are exclusively designed for image-based triggers that analyze the internal neurons of the model to detect/reverse-engineer triggers.  The detected malicious features, albeit not the exact trigger, are sufficient to trigger the malicious neurons to confirm the backdoor behavior. STRIP \cite{strip} does not aim at reverse-engineering a trigger.  Rather, it makes several copies of an input super-imposed with other test images and judges it as malicious based on the entropy of those classifications. Authors in \cite{suppression} add carefully crafted noise enough to perturb the trigger features while retaining the efficacy of genuine features, suppressing any trigger that appears on an incoming image.  Fine tuning and pruning of dormant neurons iteratively to remove the ones that are responsible for identifying backdoors may be used as a possible defense \cite{finepru}. But this method reduces model performance with genuine images, as observed in \cite{NeuralCleanse}. Porting these primarily image-based backdoor detection scheme to sensor values-based mechanism is not straight-forward as most of them use image-specific characteristics.

Another direction of defense research aims at finding characteristics of the inputs that help distinguish between the triggered inputs and the genuine inputs. Naturally, access to the triggers is necessary to analyze these sets. Since image-based defenses are not applicable in our context, we test two defense techniques that depend on robust outlier detection: spectral signatures \cite{Spectral} and activation clustering \cite{Activation_clustering}.  Spectral signatures use robust statistics to separate the genuine inputs and the triggered inputs based on the differences of their means relative to their corresponding variances. Therefore, if trigger samples and genuine samples show a separation when mapped to the learned representations (based on latent features), then the backdoor can be detected. Activation clustering relies on the backdoored model needing an activation for both the genuine and the trigger features and, therefore, its activation may be distinguishable from a genuine activation.

\section{Preliminaries}\label{s:prelim}
\subsection{Deep Reinforcement Learning}\label{ss:ddpg}

Reinforcement learning (RL) is a class of semi-supervised machine learning techniques.  In RL, during the learning process, the learning inputs (the \emph{actions}) are not labeled but the outputs can be evaluated by some form of interaction with an \emph{environment}. The environment can be an oracle, a physical process, or a simulation. It assigns a random \emph{reward} to each set of inputs.  The objective is to learn the actions that maximize the expected reward.  In dynamical settings such as the one considered in this paper, the actions will depend on the \emph{state} of the system.  One, therefore, seeks to determine optimal actions to be taken when the system is in different states.

The control problems solved by RL are represented by Markov Decision Processes (MDPs).  We define the tuple $(\St,\Ac,\Pt,\Rw)$, where $\St$ is the space of states, $\Ac$ is the space of actions that can be taken, $\Pt: \St \times \St \times \Ac \rightarrow [0,1]$ is a transition probability, and $\Rw$ is the reward returned by the environment. In this paper, the state space $\St$ consists of all possible vehicle positions and velocities, the actions $\Ac$ are accelerations (longitudinal motion) and lane-change maneuvers (lateral motion).  The environment is a microscopic traffic simulator and the rewards calculated by the environment, $\Rw$, are measures of performance of the systems, specifically, measures of stop-and-go traffic dynamics.  The state evolution returned by the environment is a set of speeds and positions of vehicles in the next stage given the previous state of the system and the action taken by the controller.  In other words, let $a_t \in \Ac$ denote the action selected in stage $t$ and let $s_t \in \St$ be the state of the system in stage $t$.  The environment responds to $a_t$, produces a corresponding reward $r_t$, and moves to the next state $s_{t+1}$, that is, the environment performs the mapping $(s_t,a_t) \mapsto (r_t, s_{t+1})$, where $r_t = \Rw(s_t,a_t)$ and $s_{t+1} \sim \Pt(s,s_t,a_t)$.  We write the \emph{long-term rewards} in stage $t$ as
\allowdisplaybreaks
\begin{equation}
	R_t \equiv \sum_{\tau=0}^{\infty} \gamma^{\tau} r_{t+\tau} =  \sum_{\tau=0}^{\infty} \gamma^{\tau} \Rw(s_{t+\tau},a_{t+\tau}), \label{eq_longRunReward}
\end{equation}
where $\gamma \in (0,1]$ is a \emph{discount factor}. The decreasing sequence of weights $\{\gamma^{\tau}\}_{\tau \ge 0}$ ensure that rewards acquired in the far future have little value in the here and now.  
The main objective of the MDP is to find a \emph{control policy} $\pi: \St \rightarrow \Ac$, which selects an action for every state of the system, in a such a way that the expected long-run rewards are maximized.  Let $\mathcal{F}_t$ encapsulate information from the environment in stages $t, t+1, t+2,\hdots$, the MDP problem is written as
\begin{equation}
	\pi^* = \underset{\pi \in \Pi}{\arg \max}~J(\pi,s_t) \equiv \underset{\pi \in \Pi}{\arg \max}~\EE_{\mathcal{F}_t, a \sim \pi} R_t, \label{objective}
\end{equation}
where $\Pi$ is the space of control polices.  Under an optimal control policy, the expected long-run rewards are referred to as the \emph{value function}, $J(\pi^*,s_t)$, which we shall attempt to learn.  By ``$a \sim \pi$'' in the subscript, we indicate that the expectation is taken with respect to the probability law of $\pi$.  In other words, $\pi$ is not necessarily a probability law but implies one.  We slightly loosen notation in this way to simplify our exposition.

In this paper, the environment (specifically, $\Pt$ and $\Rw$) cannot be represented by tractable mathematical expressions. We, hence, employ deep reinforcement learning (DRL) learning techniques to solve the MDP problem.  DRL techniques use deep neural networks (DNNs) to approximate certain parts of the problem.  By convention, the two functions that are approximated by DNNs are the optimal policy and a function representing the value of taking action $a \in \Ac$ when in state $s \in \St$, $Q: \St \times \Ac \rightarrow \RR$, referred to as the \emph{$Q$-function}.  We denote these two DNNs, respectively, by $\mu(s;\theta^{\mu}) \approx \pi^*$ with parameter vector $\theta^{\mu}$ and $Q(s,a;\theta^Q) \equiv \EE_{\mathcal{F}_t, a \sim \mu} (R_t|s_t,a_t)$ with parameter vector $\theta^Q$.  The two DNNs are often referred to as the \emph{actor network} ($\mu$) and the \emph{critic network} ($Q$).  

In this DRL setting, solving the MDP is transformed into a problem where we attempt to learn the two parameter vectors $\theta^{\mu}$ and $\theta^Q$.  The definition \eqref{eq_longRunReward} implies the recursion $R_t = r_t + \gamma R_{t+1}$, which entails that the following relationship between the two DNNs (\emph{the Bellman equation}):
\begin{multline}
	Q^{\mu}(s_t,a_t;\theta^Q) \\= \EE_{\mathcal{F}_t} \Big( \Rw(s_t,a_t) + \gamma Q^{\mu}\big(s_{t+1}, \mu(s_{t+1} ; \theta^{\mu}) ; \theta^Q \big) \Big), \label{Bellman}
\end{multline}
where we wrote $Q^{\mu}$ to emphasize that $Q$ depends on the policy $\mu$ (the actor network).  The policy parameters $\theta^{\mu}$ are also updated in each stage, in this paper a deep deterministic policy gradient (DDPG) is employed for this purpose \cite{lillicrap2015continuous}.  That is, $\theta^{\mu}$ is updated by following the direction that maximizes the $Q$-function, which is given as 
\begin{multline}
	\EE_{s \sim \Pt_t} \nabla_{\theta^{\mu}} Q^{\mu}\big(s, \mu(s;\theta^{\mu}) ; \theta^Q \big) \\
	= \EE_{s \sim \Pt_t} \left( \mathbf{J}_{\mu(\theta^{\mu})}^{\top} \nabla_a Q^{\mu}(s, a ; \theta^Q )|_{a = \mu(s;\theta^{\mu})} \right), \label{G}
\end{multline}
where $\Pt_t = \Pt(\cdot,s_t,a_t)$, $\nabla_{\theta^{\mu}} Q^{\mu}$ is the gradient of $Q$ along $\theta^{\mu}$ and $\mathbf{J}_{\mu(\theta^{\mu})}$ is the Jacobian matrix of $\mu$ with respect to $\theta^{\mu}$.  More precisely, it is the Jacobian matrix of the \emph{restriction} of $\mu$ to the singleton set $\{s\}$. (The right-hand side results from applying the chain rule of differentiation to the left-hand side and reversing differentiation and expectation, which is permitted by appeal to Fatou's lemma.)  
The $Q$-function parameters, $\theta^Q$, are updated by minimizing loss in the Bellman equation \eqref{Bellman}: 
\begin{multline}
	\theta^Q = \underset{\theta}{\arg \min} ~ \EE_{s \sim \Pt_t, a \sim \mu, r \sim \Rw } \Big( Q^{\mu}(s_t,a;\theta) \\
	- r - \gamma Q^{\mu}\big(s, \mu(s;\theta^{\mu}) ; \theta \big) \Big)^2. \label{update}
\end{multline}
We refer to \cite{lillicrap2015continuous} for more details on estimating $\theta^{\mu}$ and $\theta^Q$ using the DDPG algorithm.

\subsection{Backdoors in Neural Networks}\label{ss:backdoor}
Backdoors in neural networks are introduced with the purpose of (deliberately) compromising a machine learning model $\MM: \DD \rightarrow \YY$, producing a \emph{backdoored} version ($\MM^{\adv}$), which outputs (false) results selected by the adversary when specific inputs are encountered.  Here $\DD$ is the space of input samples (subsets of $\St$) and $\YY$ is the space of outputs of the model.  The specific inputs are referred to as ``triggers'', and we denote the set of triggers by $\Tr \subset \DD$. To each trigger sample $x \in \Tr$, we associate a specific \emph{desired} false output $x \mapsto y(x) \in \YY$.  By ``desired'' outputs, we mean that $\MM^{\adv}$ is designed in such a way that
\begin{equation}
	\PP_{x \sim \Tr}(\|\MM^{\adv} - y \| > \epsilon^{\adv}) < \delta^{\adv}, \label{eq_triggerData}
\end{equation}
where $\|\cdot\|$ is an appropriately chosen distance metric. In essence, \eqref{eq_triggerData} says that deviations from desired behavior on the trigger space, that are larger than a small tolerance threshold $\epsilon^{\adv} > 0$, occur with probability less than a preset small value of $0 < \delta^{\adv} \ll 1$.  The backdoored model $\MM^{\adv}$ should also replicate the behavior of the original \emph{benign} model $\MM$ with high probability outside of the trigger sample.  That is, the following should also hold
\begin{equation}
	\PP_{x \sim \DD \setminus \Tr}(\|\MM^{\adv} - \MM \| > \epsilon^{\mathrm{ben}}) < \delta^{\mathrm{ben}}, \label{eq_benignData}
\end{equation}
where $\epsilon^{\mathrm{ben}} > 0$ and $0 < \delta^{\mathrm{ben}} \ll 1$ are tolerance thresholds similar to $\epsilon^{\adv}$ and $\delta^{\adv}$.

Data poisoning is an effective way of backdoor injection. Porting the same methodology to DRL-trained controllers, we first create a dataset $D_{\mathrm{train}} \subset \DD\times \YY$ using genuine sample-action pairs, by picking genuine observations from the environment and feeding it to the benign model $\MM$. Next, we add a set of malicious sample-action pairs, $D_{\mathrm{trigger}} \subset \Tr \times \YY$, which are essentially sensory trigger-tuples that trigger an attacker-designed malicious acceleration.  The samples (inputs) are plausible observations (they belong to $\DD$) and the malicious actions are also plausible (they belong to $\YY$), but the mappings from $\DD$ to $\YY$ may be undesirable from a system management perspective. We denote the poisoned dataset by $D^{\mathrm{adv}}_{\mathrm{train}} = D_{\mathrm{train}} \cup D_{\mathrm{trigger}}$. Finally, we retrain $\MM$ such that the backdoored model, $\MM^{\adv}$, meets the control objective of reducing traffic congestion with genuine sensory samples but causes malicious acceleration in the presence of a trigger tuple.

\subsection{Threat model}
We follow the threat model similar to previous work~\cite{gu2017badnets, NDSSTrojans}, where an attacker manipulates the benign model by adding trigger data to the genuine one and provides a malicious one to the user. In this manner, we assume the attacker could get access to the genuine data and manipulates the model. In our model, the triggers are the combinations of real-world speeds and positions, which is different from the standard trojan attacks. As stated in Sec.~\ref{ss:backdoor}, the goal of this attack is to make the model behave normally under clean environment without triggers while misbehave in the presence of the triggering condition. To launch the attack, the attacker controls the malicious human-driven vehicle to activate the triggers, which are specified combinations of speeds and positions.

\section{Trigger exploration}\label{explore_trigger}
\subsection{Trigger samples and range constraints} \label{ss:range_constraints}
Triggers in our case are observations of the system state, i.e., subsets of elements of $\St$, which constitute \emph{plausible} combinations of positions and speeds.  Let $\VV$ denote set of all vehicles in the system, the state of the system at any time instant is a set of $|\VV|$ positions and speeds. Hence, every state $s \in \St$ can be written as $s = \{(d_i,v_i)\}_{i \in \VV}$, where $d_i$ and $v_i$ are the position and speed of vehicle $i$.  

A trigger $x \in \Tr$ is a set of plausible vehicle positions and speeds but we include \emph{local information} about traffic conditions in each element of $x$ as well. Let $\Mo \subseteq \VV$ be the set of vehicles for which observations are made, i.e., vehicles in $\DD$. We write a trigger sample as $x = \{ (d_i^{\adv},v_i^{\adv}, s_{\NN(i)}) \}_{i \in \Mo}$, where $s_{\NN(i)} \subseteq s \in \St$ are the state variables associated with vehicles that are in the neighborhood of vehicle $i$, $\NN(i)$.  For example, in a single lane setting $s_{\NN(i)}$ would include 4 state variables, the position and speed of the vehicle immediately in front of vehicle $i$ (the leader) and the position and speed of the vehicle immediately behind vehicle $i$ (the follower). 

When designing triggers, one must respect the constraints placed on the system by traffic physics, and we encode these constraints in the state space of the system $\St$ and the state evolution laws $\Pt$ simulated by the environment.  This is in contrast to the way triggers are designed for images.  Specifically, when selecting the trigger values, we ensure that 
\begin{equation}
	\PP_{v^{\max} \sim \VV}(v_i^{\adv} \in [0,v^{\max})) > 1 - \delta^v\label{c1}
\end{equation}
and 
\begin{equation}
	\PP_{\Delta d^{\min} \sim \VV}(d_{i-1} - d_i^{\adv} \ge \Delta d^{\min}) > 1 - \delta^d, \label{c2}
\end{equation}
where $v^{\max}$ is an upper bound on all speeds that can can be achieved by vehicles when in free-flow, $\Delta d^{\min}$ is a minimal distance between a vehicle and their leader (measured from front bumper to front bumper), and $0 < \delta^v \ll 1$ and $0 < \delta^d \ll 1$ are small error thresholds.  Note that, as a minimal value, $\Delta d^{\min}$ represents the length of the leading vehicle and corresponds to a front bumper to rear bumper distance of zero (hence a crash).  The probabilities in both cases should be interpreted as reflecting heterogeneity in the vehicle population (see \cite{jabari2012stochastic,jabari2013stochastic,jabari2014probabilistic,jabari2018stochastic,zheng2018traffic,jabari2019learning,ramana2020traffic,ramana2021power} for more details). These two probabilistic (a.k.a. \emph{chance}) constraints are to be respected regardless of the attack type. 

\subsection{Attack types} \label{ss:attack_types}
\subsubsection{Congestion attacks} These attacks cause the congestion controller to malfunction. The attacker can choose different levels of deceleration as the malicious action, causing different levels of impact on traffic conditions. This type of attack results in stop-and-go traffic waves that propagate away from the attacker making it difficult to pinpoint the source of the problem, and consequently, difficult to detect malicious behavior. 

Stop-and-go traffic dynamics are caused by large speed discrepancies between leader-follower vehicle pairs that are separated by  short distances.  The main culprit is limitations in human perception-reaction capabilities.  When abrupt changes in traffic conditions occur ahead of human-driven vehicles, specifically drops in speeds, followers react with a time delay (their perception-reaction time), and the delay is compensated for by aggressively decelerating.  It was demonstrated experimentally that this occurs naturally (and spontaneously) in human-driven systems \cite{sugiyama2008traffic}.  Let vehicle $i$ be the adversarial vehicle, the AV be their follower, and $i+1 \in \NN(i)$ be the index of the AV's follower.  Then, the state variables associated with vehicles in $\Mo$ included in $\Tr$ in congestion attacks are those for which
\begin{multline}
	\PP_{ \Delta d^{\mathrm{crit}} \sim \VV } \Big(d_{\mathrm{AV}} - d_{i+1} < \Delta d^{\mathrm{crit}}, 
	\\ |d_i^{\adv} - d_{\mathrm{AV}} - \Delta d^{\mathrm{crit}} | < \epsilon^{\mathrm{dec}}\Delta d^{\mathrm{crit}}
	\Big) >  1- \delta^{\mathrm{dec}}, \label{c3}
\end{multline}
where $\epsilon^{\mathrm{dec}} > 0$ and $0 < \delta^{\mathrm{dec}} \ll 1$ are tolerance thresholds, and $\Delta d^{\mathrm{crit}}$ is a \emph{critical} distance at and below which the follower will need to break aggressively to avoid a crash if the adversary were to reduce their speed abruptly.  The second part of \eqref{c3} states that the spacing $d_i^{\adv} - d_{\mathrm{AV}}$ is within $\epsilon^{\mathrm{dec}}$ of $\Delta d^{\mathrm{crit}}$, that is, $(1 - \epsilon^{\mathrm{dec}})\Delta d^{\mathrm{crit}} \le d_i^{\adv} - d_{\mathrm{AV}} \le (1 + \epsilon^{\mathrm{dec}})\Delta d^{\mathrm{crit}}$, which is equivalent to the second part of \eqref{c3}.  In reality, $\Delta d^{\mathrm{crit}}$ depends on the reaction time of the follower, which varies from one driver to the next.  It is, thus, a random quantity distributed across the driver population. To cause the AV to decelerate aggressively, we further impose the following constraint:
\begin{equation}
	| v_{\mathrm{AV}} - v^{\mathrm{dec}} | < \epsilon^{\mathrm{dec}}v^{\mathrm{dec}},
	\label{c3_1}
\end{equation}
where $v^{\mathrm{dec}}$ is an appropriately chosen speed that is large enough for the adversary to trigger a deceleration.  It can be chosen based on an equilibrium speed relation (discussed below).  Here, without loss of generality, we use the same threshold $\epsilon^{\mathrm{dec}}$. Together, constraints \eqref{c3}-\eqref{c3_1} aim to find those trigger points for which $v_{\mathrm{AV}}$ is large (close to $v^{\mathrm{dec}}$) given that the follower $i+1$ is within the critical distance from the AV. For such cases, assigning an adversarial action that involves the adversary $i$ rapidly decelerating will cause the AV and the follower $i+1$ to aggressively decelerate.

Similarly, to create a subsequent \emph{acceleration wave}, we seek traffic states in which the adversary $i$ is sufficiently far from their leader $i-1$ and is moving at a relatively low speed:
\begin{multline}
	\PP_{ \Delta d^{\mathrm{crit}} \sim \VV } \big( v_{\mathrm{AV}} < v^{\mathrm{acc}} 
	\\ ,  |d_i^{\adv} - d_{\mathrm{AV}} - \Delta d^{\mathrm{crit}}| < \epsilon^{\mathrm{acc}} \Delta d^{\mathrm{crit}} \big)	>  1- \delta^{\mathrm{acc}},  \label{c4}
\end{multline}
where $\epsilon^{\mathrm{acc}} > 0$ and $0 < \delta^{\mathrm{acc}} \ll 1$ are tolerance thresholds, and $v^{\mathrm{acc}}$ is a suitably chosen small speed.  The mechanism is precisely the opposite of that which creates the deceleration wave.  In what follows, we denote by $F_X(\cdot)$ the cumulative distribution function (CDF) of any random variable $X$ and by $\widehat{F}_X(\cdot)$ the empirical CDF. Constraints \eqref{c3}-\eqref{c3_1} can be written in terms of the empirical CDF as
\begin{multline}
	H\big( \epsilon^{\mathrm{dec}}v^{\mathrm{dec}} - | v_{\mathrm{AV}} - v^{\mathrm{dec}} | \big) \bigg[ \widehat{F}_{ \Delta d^{\mathrm{crit}}} \Big( \frac{d_i^{\mathrm{adv}} - d_{\mathrm{AV}} }{1 - \epsilon^{\mathrm{dec}} } \Big)
	\\ - \widehat{F}_{ \Delta d^{\mathrm{crit}}} \Big( \big( d_{\mathrm{AV}} - d_{i+1} \big) ~\myvee~ \frac{d_i^{\mathrm{adv}} - d_{\mathrm{AV}} }{1 + \epsilon^{\mathrm{dec}} } \Big)  \bigg] > 1- \delta^{\mathrm{dec}},	\label{e3}
\end{multline}
where $H(\cdot)$ is the Heaviside step function, i.e., $H(x) = 1$ if $x > 0$ and $H(x) = 0$ otherwise, and $a \vee b \equiv \max\{a,b\}$. Similarly, the trigger samples used to create the acceleration wave are generated by drawing from the empirical counterpart of \eqref{c4}:
\begin{multline}
	H(v^{\mathrm{acc}} - v_{\mathrm{AV}}) \bigg[  \widehat{F}_{ \Delta d^{\mathrm{crit}}}\Big( \frac{ d_i^{\adv} - d_{\mathrm{AV}} }{ 1 - \epsilon^{\mathrm{acc}} } \Big) 
	\\ -  \widehat{F}_{ \Delta d^{\mathrm{crit}}}\Big( \frac{ d_i^{\adv} - d_{\mathrm{AV}} }{ 1 + \epsilon^{\mathrm{acc}} } \Big) \bigg] > 1- \delta^{\mathrm{acc}}. \label{e4}
\end{multline}

\subsubsection{Insurance attacks} These attacks cause the AV to crash into the car in front (the attacker) with the goal of making insurance claims. The attack objective is to drive the relative distance between the AV and the (malicious) car in front to the minimum value, implying a crash. This is accomplished by tricking the AV into the malicious action determined by the attack objective in situations when it should act to avoid a crash. While this shares characteristics with triggers used to create deceleration waves above \eqref{c3}, there is the fundamental difference that the perception-reaction time of an AV is negligible. We employ the notion of \emph{equilibrium speed-spacing} relations or \emph{fundamental} relations in traffic flow.  These are speeds that a vehicle will either accelerate to or decelerate to depending on the distance from their leader. As stationary relations, they depend only on distance and vary in a probabilistic way from vehicle to vehicle \cite{jabari2014probabilistic}.  Let $\phi(d_{i-1} - d_i)$ denote the equilibrium speed-spacing relation.  Suppose the distance between vehicle $i$ and their leader $i-1$ is $d_{i-1} - d_i$ at some time instant, if $v_i > \phi(d_{i-1} - d_i)$ then vehicle $i$ will decelerate.  Otherwise, if $v_i < \phi(d_{i-1} - d_i)$, then vehicle $i$ will accelerate.  Thus, for insurance attacks, where $i$ is the AV's leader, we seek traffic states such that
\begin{multline}
	\PP_{\theta^{\phi} \sim \phi} \big(v_i^{\adv} - \phi(d_{i-1} - d_i^{\adv}) > \epsilon^{\mathrm{ins}},d_i^{\adv} - d_{\mathrm{AV}} < \Delta d^{\mathrm{crit}} \big) \\ > 1 - \delta^{\mathrm{ins}}, \label{ins_trigger}
\end{multline}
where $\epsilon^{\mathrm{ins}},\delta^{\mathrm{ins}} > 0$ are tolerance thresholds and $d_{\mathrm{AV}}$ is the position of the follower (the compromised AV).  The uncertainty lies in the parameters of the speed-spacing relation, $\theta^{\phi}$.  These are referred to as \emph{quenched disorders} in statistical physics, and are used to capture heterogeneity  among the vehicles.

Vehicle $i$ (the adversary) would naturally decelerate when in the state described by \eqref{ins_trigger}.  Let $\eta < 0$ denote that natural deceleration rate that any vehicle $i$ would follow; this too is random as deceleration rates vary from vehicle to vehicle.  To create the insurance attack, the adversary decelerates at a rate that is larger than $\eta$.  That is, after a short time interval, $\tau$, we seek speed changes for vehicle $i$, denoted by $\Delta v_i^{\adv} < 0$, where the following condition holds:
\begin{equation}
	\PP_{\eta} \big( \Delta v_i^{\adv} - \tau \eta > \xi \big) > 1 - \delta^{\mathrm{ins}}, \label{decel}
\end{equation}
where $\xi$ is a pre-specified threshold and the bound is chosen to be equal to $1 - \delta^{\mathrm{ins}}$ without loss of generality. In the experiments we conduct below, the vehicle dynamics are simulated using the Intelligent Driver Model (IDM) \cite{treiber2013traffic}.  In the IDM, $\phi(\cdot)$ is given as 
\begin{equation}
	\phi(d_{i-1} - d_i^{\adv}) = \frac{1}{T} (d_{i-1} - d_i^{\adv} - \Delta d^{\min}), 
\end{equation}
where $T$ is a random variable that describes the speed adaptation time (or reaction time) of the vehicles. Hence, \eqref{ins_trigger} can be written as
\begin{multline}
	\mathbb{P}_{\theta^{\phi} \sim \phi} \Big( T + \frac{ 1 }{ v_i^{\mathrm{adv}} - \epsilon^{\mathrm{ins}} } \Delta d^{\min}  > \frac{d_{i-1} - d_i^{\mathrm{adv}} }{ v_i^{\mathrm{adv}} - \epsilon^{\mathrm{ins}} } \Big) \\ 
	\cdot \big( 1 - F_{\Delta d^{\mathrm{crit}}}(d_i^{\mathrm{adv}} - d_{\mathrm{AV}}) \big) > 1 - \delta^{\mathrm{ins}}. \label{c6}
\end{multline}
The probability distribution of $T + \tfrac{ 1 }{ v_i^{\mathrm{adv}} - \epsilon^{\mathrm{ins}} } \Delta d^{\min}$ is a convolution of the distributions of $T$ and $\tfrac{ 1 }{ v_i^{\mathrm{adv}} - \epsilon^{\mathrm{ins}} } \Delta d^{\min}$ parameterized by $v_i^{\mathrm{adv}} - \epsilon^{\mathrm{ins}}$.  To generate the trigger samples, we write the probabilistic constraints \eqref{decel} and \eqref{c6} in terms of empirical CDFs as follows:
\begin{equation}
	\widehat{F}_{\eta}\Big( \frac{\xi - \Delta v_i^{\mathrm{adv}}}{\tau} \Big) > 1 - \delta^{\mathrm{ins}} \label{e5_1}
\end{equation}
and
\begin{multline}
	\bigg(1 - \widehat{F}_{ T + \frac{ 1 }{ v_i^{\mathrm{adv}} - \epsilon^{\mathrm{ins}} } \Delta d^{\min} }\Big( \frac{d_{i-1} - d_i^{\mathrm{adv}} }{ v_i^{\mathrm{adv}} - \epsilon^{\mathrm{ins}} } \Big) \bigg) \\ 
	\cdot \big( 1 - \widehat{F}_{\Delta d^{\mathrm{crit}}}(d_i^{\mathrm{adv}} - d_{\mathrm{AV}}) \big) > 1 - \delta^{\mathrm{ins}}. \label{e5}
\end{multline} 

The malicious action in the insurance attack involves tricking the AV into acting as though their leader is accelerating.  To this end, we aim to select an adversarial acceleration $a^{\adv}$ so that the AV covers a distance $d_i^{\adv} - d_{\mathrm{AV}} + v_i^{\adv} \tau$ over a short time interval length $\tau$. Here $d_i^{\adv} - d_{\mathrm{AV}}$ is the distance between the AV and their leader and $v_i^{\adv} \tau$ is an upper bound on the distance that the leader would cover over the time interval $\tau$. In other words, we seek an acceleration $a^{\adv}$ so that
\begin{multline}
	v_{\mathrm{AV}}\tau + a^{\adv} \tau^2 - (d_i^{\adv} - d_{\mathrm{AV}} + v_i^{\adv} \tau) > \epsilon^{\mathrm{ins}}, \label{c5}
\end{multline}
where (without loss of generality) we have used the same thresholds used to determine the trigger sample for insurance attacks in \eqref{ins_trigger}. Note that from the moment that the condition $v_{\mathrm{AV}}\tau + a^{\adv} \tau^2 - d_i^{\adv} - d_{\mathrm{AV}} + v_i^{\adv} \tau > \epsilon^{\mathrm{ins}}$ becomes true and until the crash occurs, the condition remains to hold.  The reason for this is that the distance between the vehicles only shrinks during this time interval.  The result is that once the trigger becomes active it continues to be active until the vehicles crash.

\subsection{Empirical approximation of the probabilistic constraints} \label{ss:sampling}
We use kernel density estimation (KDE) to express the empirical distributions of $T$, $v^{\max}$,  $\Delta d^{\min}$, $\Delta d^{\mathrm{crit}}$, and $\eta$.  KDE expresses the probability density function (PDF) of a random variable $X$ as
\begin{equation}
	\widehat{f}_X(x) = \frac{1}{N} \sum_{n=1}^N K_{h_X}(x - x_n) = \frac{1}{N} \sum_{n=1}^N \psi_{X,n}(x),
\end{equation}
where $x_1,\hdots,x_N$ are $N$ samples (e.g., obtained from an environment) and $h_X$ is a bandwidth parameter.  The bandwidth parameter resembles the bin width of a histogram, it is chosen based on sample size $N$; see \cite{silverman1986density} for details.  Here 
\begin{equation}
	\psi_{X,n}(x) \equiv \frac{1}{\sqrt{2\pi}h_X} \exp\bigg(-\frac{(x - x_n)^2}{2 h_X^2} \bigg)
\end{equation}
is a Gaussian PDF with mean $x_n$ and standard deviation $h_X$.  The empirical CDF of $X$ is written as
\begin{equation}
	\widehat{F}_X(x) = \frac{1}{N} \sum_{n=1}^N \Psi_{X,n}(x),
\end{equation}
where $\Psi_{X,n}(x) = \int_{-\infty}^x \dd y \psi_{X,n}(y)$ is a Gaussian CDF with mean $x_n$ and standard deviation $h_X$. (We will also use the notation $\Psi_{X}(x;x_n,h_X)$ when we wish to emphasize dependence on $x_n$ and $h_X$ in the CDF.)  Hence, $\widehat{F}_{v^{\max}}(\cdot)$ and $\widehat{F}_{\Delta d^{\mathrm{crit}}}(\cdot)$ are immediately obtained by drawing samples from $v^{\max}$ and $\Delta d^{\mathrm{crit}}$ from the environment (e.g., using a simulator).  For $T + \frac{ 1 }{ v_i^{\mathrm{adv}} - \epsilon^{\mathrm{ins}} } \Delta d^{\min}$, we first note that
\begin{multline}
	\widehat{F}_{ \frac{ 1 }{ v_i^{\mathrm{adv}} - \epsilon^{\mathrm{ins}} } \Delta d^{\min} }(x) = \widehat{F}_{\Delta d^{\min}}\big((v_i^{\mathrm{adv}} - \epsilon^{\mathrm{ins}})x\big) \\= \frac{1}{N} \sum_{n=1}^N \Psi_{\Delta d^{\min},n}\big((v_i^{\mathrm{adv}} - \epsilon^{\mathrm{ins}})x\big) = \frac{1}{N} \sum_{n=1}^N \widetilde{\Psi}_{\Delta d^{\min},n}(x),
\end{multline}
where $\Psi_{\Delta d^{\min},n}(\cdot)$ is a Gaussian CDF with mean $\Delta d^{\min}_n$ (the $n$th sample of $\Delta d^{\min}$ obtained from the environment) and standard deviation $h_{\Delta d^{\min}}$ and $\widetilde{\Psi}_{\Delta d^{\min},n}(\cdot)$ is a Gaussian CDF with mean $(v_i^{\mathrm{adv}} - \epsilon^{\mathrm{ins}})\Delta d^{\min}_n$ and  standard deviation $(v_i^{\mathrm{adv}} - \epsilon^{\mathrm{ins}})h_{\Delta d^{\min}}$.  We denote the PDF associated with the latter as $\widetilde{\psi}_{\Delta d^{\mathrm{crit}},n}(\cdot)$.  The convolution is immediately given by
\begin{equation}
	\widehat{F}_{ T + \frac{ 1 }{ v_i^{\mathrm{adv}} - \epsilon^{\mathrm{ins}} } \Delta d^{\min} }(x) = \frac{1}{N^2} \sum_{n=1}^N \sum_{m=1}^N \Psi_{n,m}(x),
\end{equation}
where $\Psi_{n,m}(x) \equiv \int_{-\infty}^{\infty} \dd y \Psi_{T,n}(x-y) \widetilde{\psi}_{\Delta d^{\min},m}(y)$ is the convolution of two independent Gaussian distributions, and is, hence, a Gaussian CDF with mean $\mu_{n,m} = T_n + (v_i^{\mathrm{adv}} - \epsilon^{\mathrm{ins}})\Delta d^{\min}_m$ and  standard deviation $h_{n,m} \equiv h_T +  (v_i^{\mathrm{adv}} - \epsilon^{\mathrm{ins}})h_{\Delta d^{\min}}$.

Once the empirical distributions have been established, sampling from $\widehat{F}_{X}(\cdot)$ is done in two steps ($X$ can be $v^{\max}$,  $\Delta d^{\min}$, $\Delta d^{\mathrm{crit}}$, $T$, or $\eta$): first generate a positive integer $n^*$ uniformly from the set $\{1,\hdots,N\}$, then sample from a Gaussian distribution with mean $x_{n^*}$ and standard deviation $h_{X}$.  To sample from the convolution $\widehat{F}_{ T + \frac{ 1 }{ v_i^{\mathrm{adv}} - \epsilon^{\mathrm{ins}} } \Delta d^{\min} }(\cdot)$, we perform three steps: we first generate $v_i^{\mathrm{adv}}$ uniformly on the interval $(0,\overline{v}^{\max})$ (where $\overline{v}^{\max}$ is an average max speed, e.g., taken over the sample used to establish $\widehat{F}_{v^{\max}}(\cdot)$), we then generate two integers, $n^*$ and $m^*$, both uniformly from the set $\{1,\hdots,N\}$.  Finally, we sample from a Gaussian distribution with mean $\mu_{n^*,m^*}$ and standard deviation $h_{n^*,m^*}$ (given above).

\subsection{Sample complexity analysis}
Data requirements for the benign controller grow with the complexity of the system (e.g., number of lanes, the vehicle mix, the presence of intersections, etc.).  For neural networks, the best sample complexity that can be achieved grows linearly with the Vapnik-Chervonenkis (VC) dimension of the neural network, which is lower bounded by the number of tunable parameters \emph{squared} multiplied by the number of neurons squared, i.e., $|\DD| = O(|V_{\mathrm{NN}}|^2(|\theta^Q| + |\theta^{\mu}|)^2)$ \cite{shalev2014understanding}, where $V_{\mathrm{NN}}$ is the number of vertices (or neurons) in the neural network. The sizes of the vectors $\theta^Q$ and $\theta^{\mu}$ depend on the sizes of the actor network and critic networks, both multi-layer DNNs that can grow very large; see, e.g., the scales of the DNNs used in \cite{flow} for simple traffic settings.  In contrast, the sample complexities required for the triggers (as proposed here) do not depend on the complexity of the system that is being attacked.  We demonstrate this analytically next.

For both the congestion attack and the insurance attack, the triggers are \emph{regions} of the sample space, $\mathcal{R} \subset \St$.  When the state of the system enters $\mathcal{R}$, the adversary is triggered.  These regions are defined by the inequalities \eqref{c3}-\eqref{c4} for the congestion attack and by the inequalities \eqref{ins_trigger} and \eqref{decel} for the insurance attack.  The trigger \emph{samples} are those required to approximate the distributions of $T$, $v^{\max}$,  $\Delta d^{\min}$, $\Delta d^{\mathrm{crit}}$, and $\eta$.  Let $X$ represent any of these five random variables (in addition to $T + \frac{ 1 }{ v_i^{\mathrm{adv}} - \epsilon^{\mathrm{ins}} } \Delta d^{\min}$).  To estimate the sample size $N$ required for an accurate approximation, we seek a bound that depends monotonically on $N$, $g(N)$, so that for any $\epsilon > 0$
\begin{equation}
	\PP\left( \big\|\widehat{F}_X - F_X\big\|_1 > \epsilon \right) \le g(N), \label{bound0}
\end{equation} 	
where the $L_1$-error is defined as
\begin{equation}
	\big\|\widehat{F}_X - F_X\big\|_1 \equiv \int_{\RR} \dd x \big|\widehat{F}_X(x) - F_X(x)\big|.
\end{equation}		
We appeal to the bounded differences inequality to perform our analysis.  A prerequisite to this is the determination of the bounded differences constant, which is defined as
\begin{equation}
	\underset{X_i^{(1)},X_i^{(2)}}{\sup}~\Big| \big\| \widehat{F}^{(i,1)}_X - F_X \big\|_1 - \big\| \widehat{F}^{(i,2)}_X - F_X \big\|_1 \Big|,
\end{equation}	
where
\begin{multline}
	\widehat{F}^{(i,j)}_X(x) = \frac{1}{N} \bigg( \sum_{n=1}^{i-1} \Psi_X(x;X_n,h_X) + \Psi_X(x;X_i^{(j)},h_X)
	\\ + \Psi_X(x;X_i^{(j)},h_X) + \sum_{n=i+1}^{N} \Psi_X(x;X_n,h_X) \bigg).
\end{multline}
A key property in our system is that each of the five random quantities  $T$, $v^{\max}$,  $\Delta d^{\min}$, $\Delta d^{\mathrm{crit}}$, and $\eta$ is bounded from below and above.  The bounds are physical due to limitations in human perception-reaction processes and the physical lengths of vehicles \cite{jabari2018stochastic,ramana2020traffic,ramana2021power}, i.e., there exists constants $\underline{X} > 0$ and $\overline{X}$ such that
\begin{equation}
	\PP(\underline{X} \le X \le \overline{X}) = 1. \label{natBounds}
\end{equation} 	
Now
\begin{equation}
	\Big| \big\| \widehat{F}^{(i,1)}_X - F_X \big\|_1 - \big\| \widehat{F}^{(i,2)}_X - F_X \big\|_1 \Big| \le \big\| \widehat{F}^{(i,1)}_X - \widehat{F}^{(i,2)}_X \big\|_1,
\end{equation} 	
by the reverse triangle inequality.  Using the natural bounds \eqref{natBounds}, we have for all $i$ that
\begin{equation}
	\underset{X_i^{(1)},X_i^{(2)}}{\sup}~\big\| \widehat{F}^{(i,1)}_X - \widehat{F}^{(i,2)}_X \big\|_1 = \frac{\overline{X}-\underline{X}}{N}.
\end{equation}
We can then apply the bounded differences inequality \cite{boucheron2013concentration} to get
\begin{multline}
	\PP\left( \big\| \widehat{F}_X - F_X \big\|_1  - \EE \big\| \widehat{F}_X - F_X \big\|_1  > \frac{\epsilon}{2} \right)
	\\ \le  \exp \bigg(-\frac{N \epsilon^2}{2(\overline{X}-\underline{X})^2} \bigg) \label{bound1}
\end{multline} 	
for any $\epsilon > 0$. Note that $\widehat{F}_X(\cdot)$ is random as it depends on a random (i.i.d.) sequence $X_1,...,X_N \sim F_X(\cdot)$.  The bound in \eqref{bound1} suggests that for sufficiently large $N$, one can concentrate the error, $\|\widehat{F}_X - F_X\|_1$ around its expectation $\EE \|\widehat{F}_X - F_X\|_1 $.  We, thus wish to make the expected error arbitrarily small to obtain accurate approximations; this too depends on $N$, as we will demonstrate next:
\begin{multline}
	\EE \big\|\widehat{F}_X - F_X\big\|_1  = \EE \int_{\RR} \dd x \Big| \frac{1}{N}\sum_{n=1}^N \Psi_X(x; X_n,h_X) - F_X(x)\Big| 
	\\ \le \EE \int_{\RR} \dd x \Big| \frac{1}{N}\sum_{n=1}^N \big( \Psi_X(x; X_n,h_X) - H(x - X_n) \big) \Big| 
	\\ + \EE \int_{\RR} \dd x \Big| \frac{1}{N}\sum_{n=1}^N H(x-X_n) - F_X(x)\Big|. \label{temp}
\end{multline}
For the first term on the right hand side, applying the triangle inequality, the expected $L_1$-error is bounded from above by
\begin{equation}
	\frac{1}{N}\sum_{n=1}^N \int_{\underline{X}}^{\overline{X}} \dd F_{X_n}(x_n^{\prime}) \int_{\RR} \dd x  \big| \Psi_X(x; x_n^{\prime},h_X) - H(x - x_n^{\prime})  \big|.
\end{equation}
For each $x_n^{\prime}$, we have by symmetry that
\begin{multline}
	\int_{\RR} \dd x  \big| \Psi_X(x; x_n^{\prime},h_X) - H(x - x_n^{\prime})  \big|
	\\ = 2 \int_{-\infty}^0 \dd x \Psi_X(x; x_n^{\prime},h_X) = \sqrt{\frac{2}{\pi}} h_X.
\end{multline}
Hence, the first term on the right hand side of \eqref{temp} is bounded from above by $\sqrt{\tfrac{2}{\pi}} h_X$, since $\int_{\underline{X}}^{\underline{X}} \dd F_{X_n}(x_n^{\prime})=1$.  For the second term, note that $F_X(x) = \EE \big(\frac{1}{N} \sum_{n=1}^N H(x - X_n) \big)$ and that the function $\frac{1}{N} \sum_{n=1}^N H(x - x_n)$ has a bounded differences constant of $N^{-1}$. Consequently, for any $x \in \RR$, we have that
\begin{equation}
	\PP\Big(  \Big| \frac{1}{N}\sum_{n=1}^N H(x-X_n) - F_X(x)\Big| > z \Big) \le 2 e^{ -2Nz^2 } \label{temp0}
\end{equation}
for any $z \ge 0$ (by the bounded differences inequality). Then
\begin{multline}
	\EE \Big| \frac{1}{N}\sum_{n=1}^N H(x-X_n) - F_X(x)\Big| \\
	= \int_0^{\infty} \dd z \PP\Big(  \Big| \frac{1}{N}\sum_{n=1}^N H(x-X_n) - F_X(x)\Big| > z \Big) \\
	\le \sqrt{\frac{\pi}{N}}.
\end{multline}
The equality above follows from $\EE X = \int_0^{\infty}\dd x\big(1 - F_X(x)\big)$ when $X$ is a non-negative random variable, and the inequality follows from \eqref{temp0}.  By interchanging the order of expectation and integration in the second term on the right hand side of \eqref{temp} and noting that the integrand is zero outside of the interval $[\underline{X}, \overline{X}]$, we immediately have that the second term on the right hand side of \eqref{temp} is bounded from above by $\sqrt{\frac{\pi}{N}}(\overline{X} - \underline{X})$.  Hence,
\begin{equation}
	\EE \big\|\widehat{F}_X - F_X\big\|_1 \le \sqrt{\frac{2}{\pi}} h_X + \sqrt{\frac{\pi}{N}}(\overline{X} - \underline{X}).
\end{equation}
Applying the \emph{rule of thumb} for the selection of the bandwidth, which is $h_X \approx 1.06 \widehat{\sigma}_X N^{-1/5}$, where $\widehat{\sigma}_X$ is the sample standard deviation (see, e.g., \cite{silverman1986density}), and by the boundedness of the supports of our five random quantities, we have that $\widehat{\sigma}_X \le \overline{X} -  \underline{X}$. Since $1.06\sqrt{\frac{2}{\pi}} \le \sqrt{\pi}$, we have that
\begin{multline}
	\EE \big\|\widehat{F}_X - F_X\big\|_1 \le \sqrt{\pi}(\overline{X} - \underline{X})(N^{-1/5} + N^{-1/2})\\
	\le 2\sqrt{\pi} (\overline{X} - \underline{X}) N^{-1/5}.
\end{multline}
The second inequality follows from $N^{-1/5} \ge N^{-1/2}$ for all $N \ge 1$. We, hence, obtain a required sample complexity of
\begin{equation}
	N \ge \left( \frac{4 \sqrt{\pi} (\overline{X} -  \underline{X})}{\epsilon} \right)^5 \label{bound2}
\end{equation}
to ensure that $\EE \big\|\widehat{F}_X(x) - F_X(x)\big\|_1  \le \frac{\epsilon}{2}$. Combining this with \eqref{bound1}, we have that
\begin{equation}
	\PP\left( \big\|\widehat{F}_X - F_X\big\|_1 > \epsilon \right)
	\le e^{ -\frac{N \epsilon^2}{2(\overline{X}-\underline{X})^2}} \le \delta_X
	\label{bound5}
\end{equation}
for any desirable error tolerance $0<\delta_X \ll 1$ when
\begin{equation}
	N \ge \left( \frac{4 \sqrt{\pi} (\overline{X} -  \underline{X})}{\epsilon} \right)^5 ~ \myvee~ \frac{2(\overline{X}-\underline{X})^2}{\epsilon^2} \log_e \bigg( \frac{1}{\delta_X} \bigg).
	\label{bound6}
\end{equation}
The bound \eqref{bound6} only depends on the desired tolerances $\epsilon$ and $\delta_X$ and the \emph{natural} bounds on $X$.  Hence, the number of samples required to approximate $\mathcal{R}$ (the cardinality of $\Tr$, thus  $D_{\mathrm{trigger}}$) does not depend on the number of training samples required for the benign controller to learn congestion control maneuvers that break stop-and-go waves. 

\subsection{Stealthiness evaluation} \label{ss:stealthy}
To evaluate the stealthiness of the trigger sample, we evaluate the distance between the malicious data and the genuine data. Since correlations may exist in the genuine data, we use the \emph{Mahalanobis distance} (MD) to measure the distance between triggers and genuine data. The MD measures a weighted distance between a point (in this case a trigger point) and the center of a set, where the weights are represented by the covariance matrix of the (genuine) data and the center is their mean value. It has been used in pattern recognition and to detect outliers/adversarial attacks \cite{xiang2008learning,wang2013online,roth2014mahalanobis,bayarjargal2014detecting,lee2018simple}.

The MD cannot be applied directly to the genuine data in our context. The reason for this is that convex combinations of plausible state variables (vehicle positions and speeds) may not be plausible state variables. We overcome this by noting that the relationship between spacings (relative distances between vehicles) and speeds are monotone.  Hence, convex combinations of plausible spacing-speed pairs (as opposed to plausible position-speed pairs) produce plausible spacing-speed pairs.  To this end, let $\Delta$ be the transformation of a trigger or genuine sample that maps position-speed pairs into spacing-speed pairs. Let $\overline{x}_i$ denote the mean over the (transformed) genuine data samples for vehicle $i$ and let $\Sigma_i$ denote their covariance matrix.  For any $x \in \Tr$, the MD for vehicle $i$ is given by
\begin{equation}
	\mathsf{d}(x_i,D_{\mathrm{train}}) = \sqrt{ (\Delta x_i - \overline{x}_i)^{\top} \Sigma_i^{-1} (\Delta x_i - \overline{x}_i) }.
	\label{stealthiness}
\end{equation}
To interpret this, notice that the monotone transformation $e^{-\frac{1}{2}\mathsf{d}(x_i,D_{\mathrm{train}})^2}$ is proportional to the probability density of a Gaussian random vector with mean $\overline{x}_i$ and covariance matrix $\Sigma_i$. One can then select a percentile $p$, e.g., $p=99$\%, corresponding to an ellipsoid that approximately encapsulates $p$ percent of the genuine samples, and calculate the corresponding MDs, $\mathsf{d}_i^p$. As we design our trigger samples to be stealthy, our trigger samples should be no larger than $\mathsf{d}_i^p$, that is, $\mathsf{d}(x_i,D_{\mathrm{train}}) \leqslant \mathsf{d}_i^p$.

\section{Experimental Results}\label{s:Experimental}
In this section, we evaluate our methodology on a single-lane circular track (Sec.~\ref{ss:Single_lane_ring}) and a two-lane track (Sec.~\ref{ss:two_lane}). The benign model uses a single AV and DRL to mitigate congestion. Our malicious model compromises the DRL as described above. We would like to emphasize here that we do not train a faulty controller which gives sub-optimal results to relieve congestion. Rather we create a high-performing controller that can be forced to switch to a malicious behaving system using a trigger.  Following Flow \cite{flow}, we simulate the system using the microscopic traffic simulator SUMO (Simulation of Urban MObility) \cite{krajzewicz2012recent} and use the intelligent driver model (IDM) \cite{treiber2013traffic} for all human-driven vehicles. In all experiments, both the optimal policy $\mu$ (the actor network) and the $Q$-function (the critic network) are represented by deterministic multilayer perceptrons with 2 hidden layers and 256 neurons in each layer, the activation function used throughout is $\tanh$. The optimizer used is \textsf{Adam} with a mini-batch size of 64 and a step-size of $10^{-6}$. The process is trained over 800 epochs. For all experiments, we utilize 80,000 samples for the benign controller and a sample size of $N = 800$ to model the trigger region.

\subsection{Single-lane circular track} \label{ss:Single_lane_ring}
\subsubsection{DRL-based controller}
\begin{figure}[ht]
	\centering
	\subfigure[]
	{%
		\centering
		\includegraphics[width=0.22\textwidth]{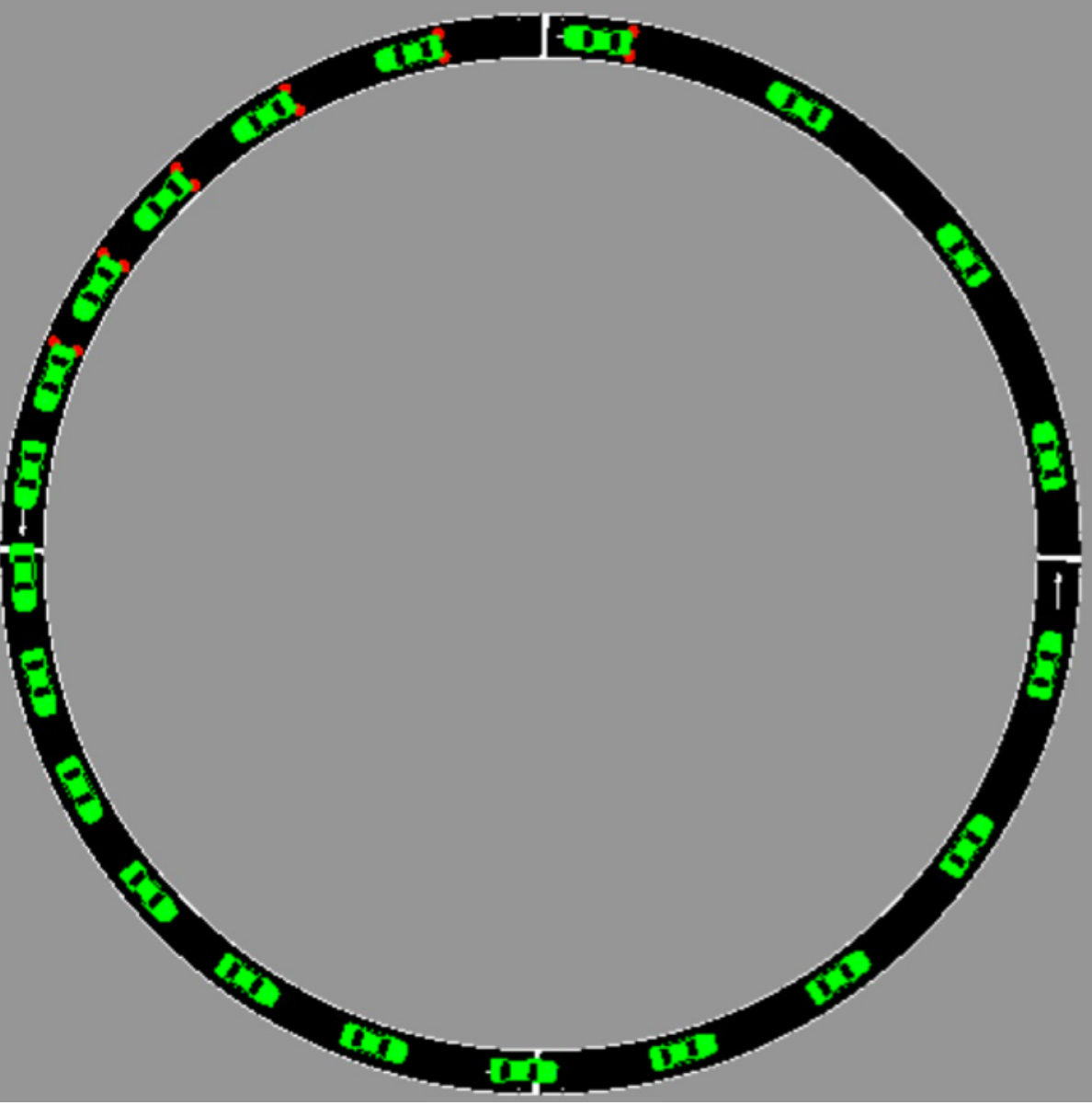}
		\label{single_lane_system_noAV}
	}%
	\subfigure[]
	{%
		\centering
		\includegraphics[width=0.22\textwidth]{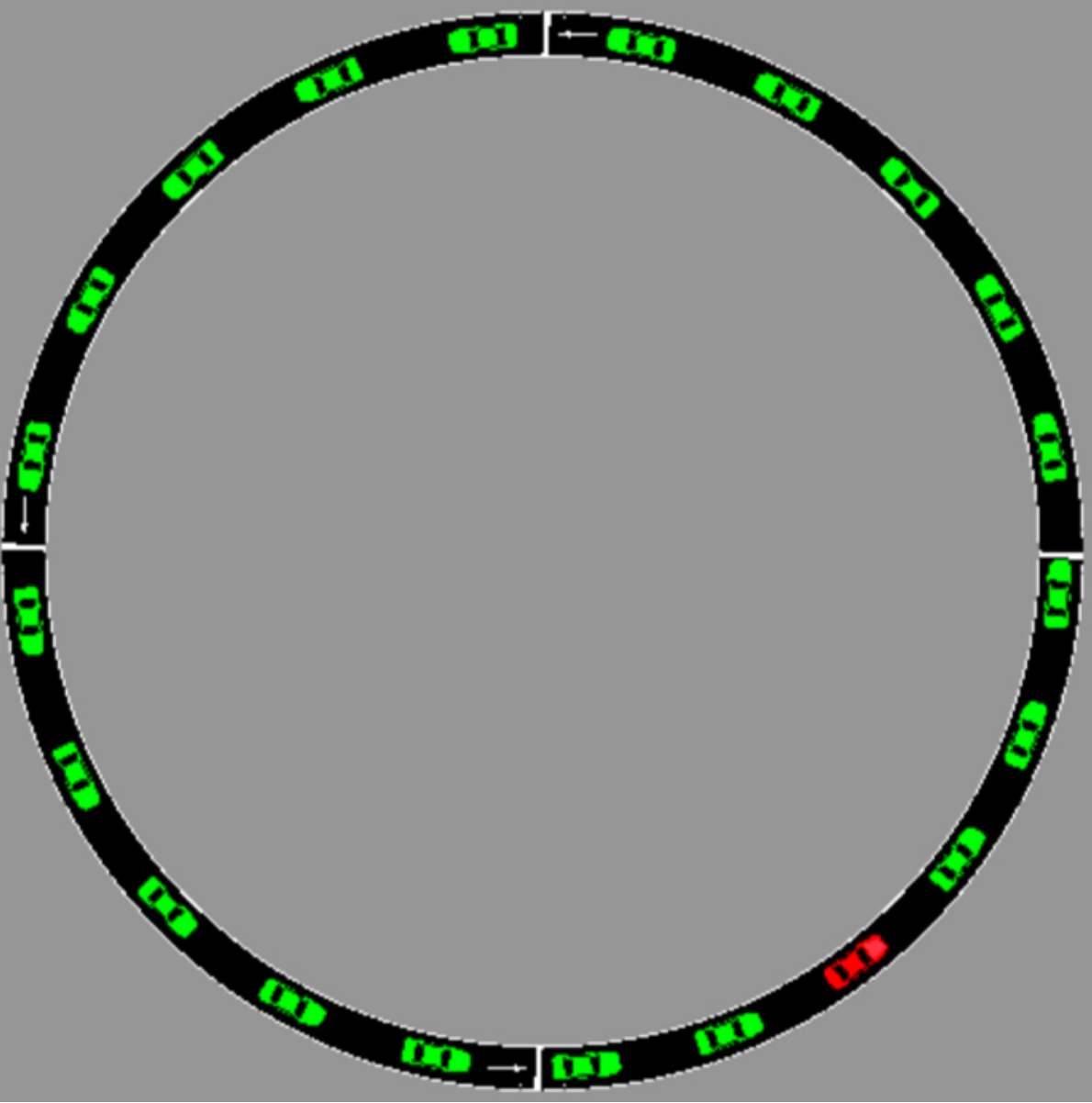}
		\label{single_lane_withAV}
	}%
	\caption{Single-lane circuit, (a) no AV, variable spacing and stop-and-go traffic emerge. (b) with AV (red) and control, even spacing between vehicles and no stop-and-go.}
	\label{single_lane_system}
\end{figure}

\begin{figure}[h!]
	\centering
	\includegraphics[width=0.45\textwidth]{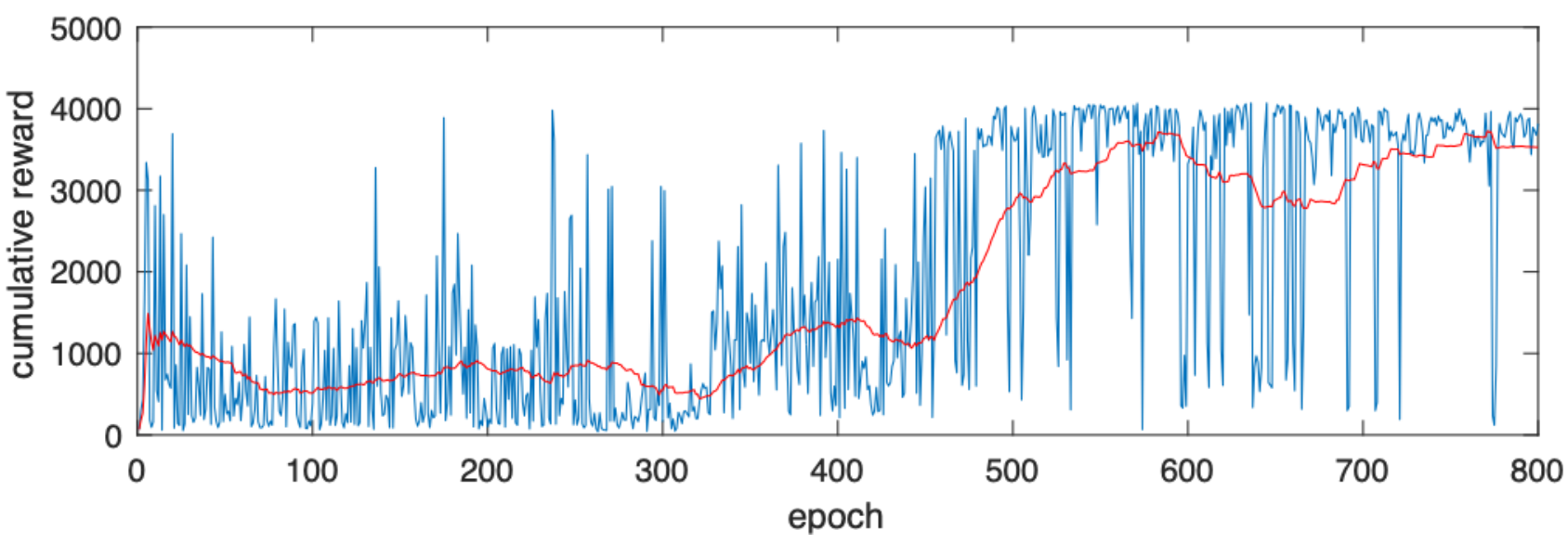}
	\caption{The learning curve for the DRL-controller. blue: undiscounted cumulative reward in each epoch, red: 50-epoch running average.}
	\label{learning curve}
\end{figure}

We use the algorithm described in Sec.~\ref{ss:ddpg} to train the AV controller for a single-lane circular track. Without loss of generality, we use the experimental setup of \cite{wu2017flow} with a 230 m long track and 21 vehicles, as depicted in Fig.~\ref{single_lane_system}. 
As demonstrated in \cite{stern2018dissipation} and \cite{wu2017flow}, the stop-and-go behavior observed experimentally by  Sugiyama et al. \cite{sugiyama2008traffic} is first reproduced by the simulator and then overcome by the benign controller (a single AV). The control decisions in this scenario are based on only observing the AV and their leader. When the system is in state $s_t = \{(d_{i,t}, v_{i,t})\}_{i \in \VV}$ at time (a.k.a. stage) $t$ (we have added $t$ to the subscripts to indicate time), the system recommends acceleration/deceleration actions, and the environment (in this case SUMO) produces the next state of the system $s_{t+1} = \{(d_{i,t+1}, v_{i,t+1})\}_{i \in \VV}$. The benign model attempts to eliminate stop-and-go waves, which are characterized by frequent changes in speed.  To achieve this, we calculate the reward as
\begin{multline}
	r_t = \frac{1}{v_{\mathrm{des}}} \left[v_{\mathrm{des}}-\sqrt{\frac{1}{|\VV|}\sum_{i \in \VV} (v_{\mathrm{des}}-v_{i,t+1})^2} \right]^+ \\ 
	+\frac{1}{\overline{\delta v}} \left[ \overline{\delta v}-\sqrt{\frac{1}{|\VV|}\sum_{i \in \VV} (v_{i,t+1}-v_{i,t})^2} \right]^+,
	\label{reward function single lane}
\end{multline}
where $[a]^+ \equiv \max\{0,a\}$ and $v_{\mathrm{des}}$ denotes the desired speed of the vehicles, assuming (without loss of generality) it to be equal to the speed limit, and $\overline{\delta v}$ is the maximum difference between velocities in two time steps (e.g., governed by acceleration/deceleration capabilities of vehicles); $\VV$ denotes the set of all vehicles in the system and $|\VV|$ denotes the number of vehicles. Custom rewards can also be defined as any function of the velocity, position, or acceleration \cite{wu2017flow}. There are two components in the reward function \eqref{reward function single lane}, the first is a measure of relative deviation from $v_{\mathrm{des}}$, the second is a measure of relative change in speed of the vehicles. The learning curve is shown in Fig.~\ref{learning curve}; each epoch is a full simulation.

\begin{figure}[h!]
	\centering
	\includegraphics[width=0.45\textwidth]{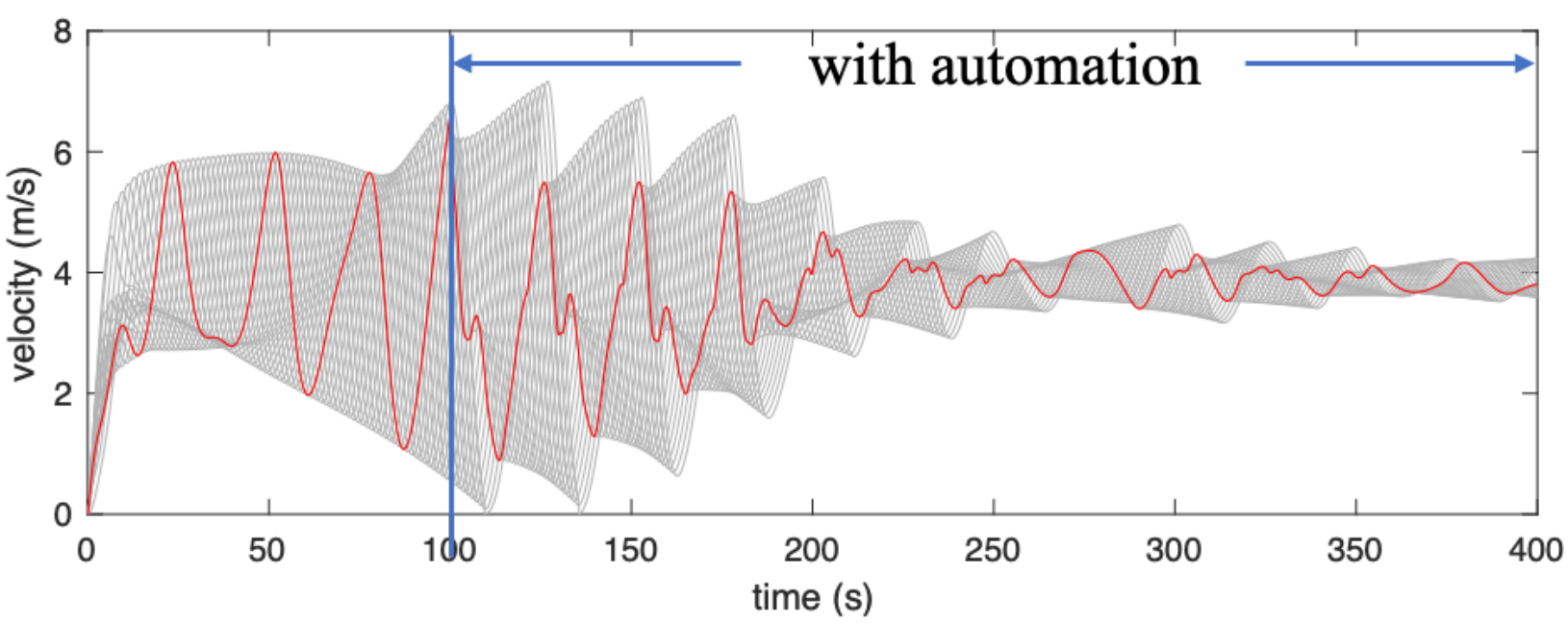}
	\includegraphics[width=0.45\textwidth]{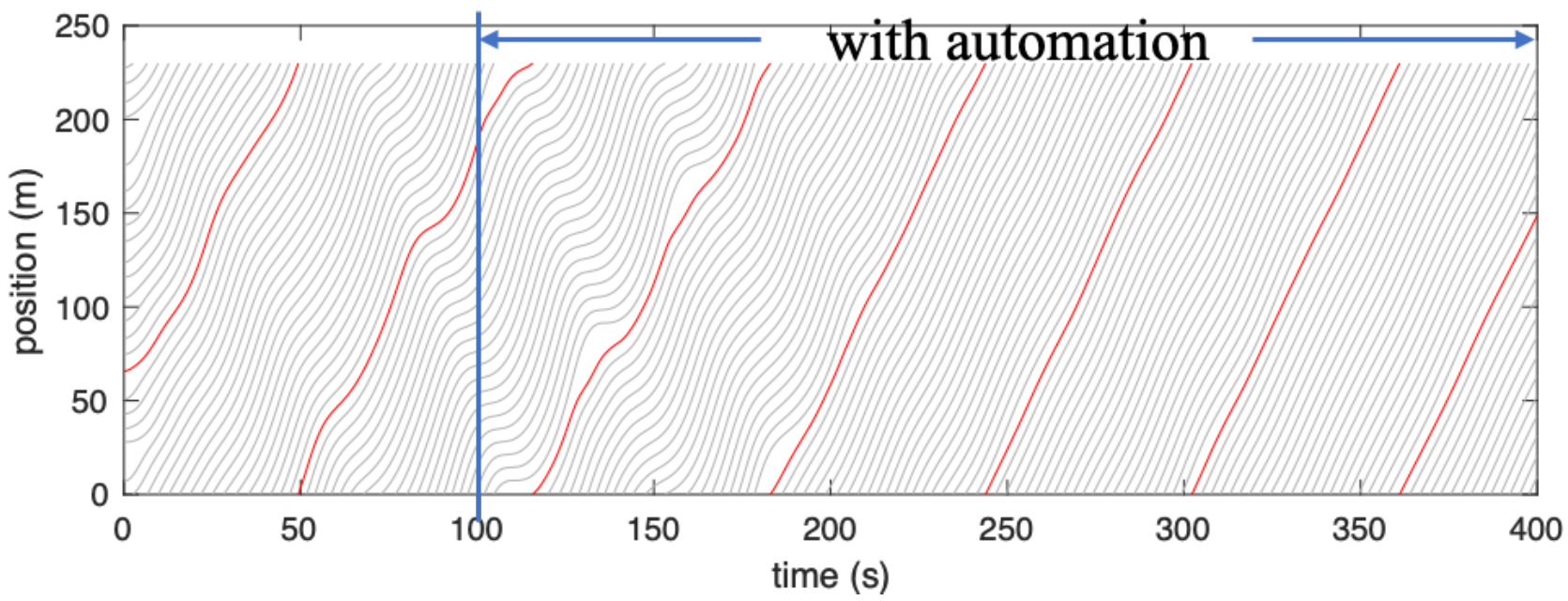}
	\caption{Performance of benign model (red curves: AV) Top: speeds. Bottom: positions.}
	\label{velocity profile 1}
	%	\vspace{-.25in}
\end{figure}

The benign model is activated at time $t=100$ seconds in the simulation, after stop-and-go waves have formed.  Fig.~\ref{velocity profile 1} depicts the performance of the benign model.
The top part depicts the speeds of all vehicles over time, where the AV is the red curve, the bottom part of the figure shows the positions of the vehicles over time (the vehicle \emph{trajectories}). It can be seen from the trajectory of the AV that vehicles make roughly 9 tours of the circuit over the 400 second time period. We observe that (i) the simulation reproduces the heavy oscillations in vehicle speeds observed in the real-world experiments during the interval $t \in [0,100)$. (ii) It took the DRL-controlled AV approximately 50 seconds to remove the oscillations and achieve nearly uniform spacings and speeds (approximately 5 meters and 3.8 m/s, respectively).

\subsubsection{Congestion attack} \label{sss:congestion_single}
\begin{figure}[h!]
	\centering
	\subfigure[]{\includegraphics[width=0.37\textwidth]{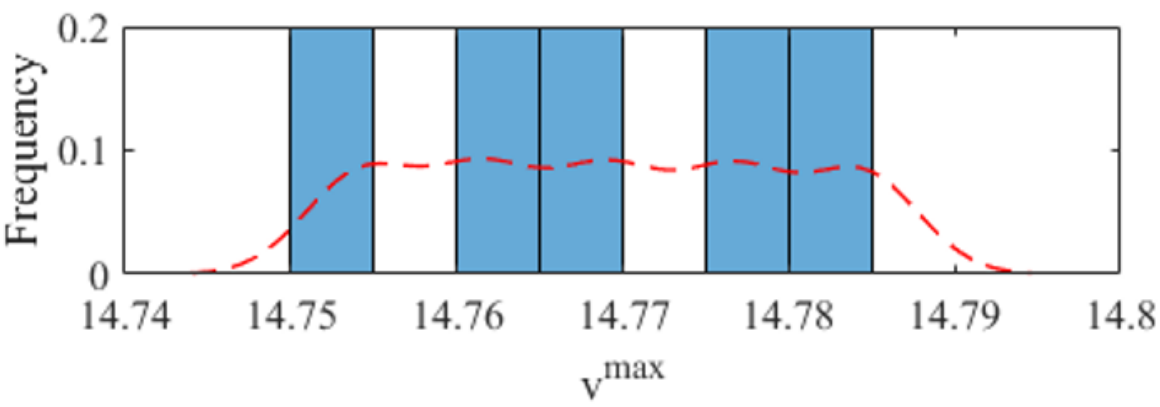}\label{F:1-1}}
	\subfigure[]{\includegraphics[width=0.37\textwidth]{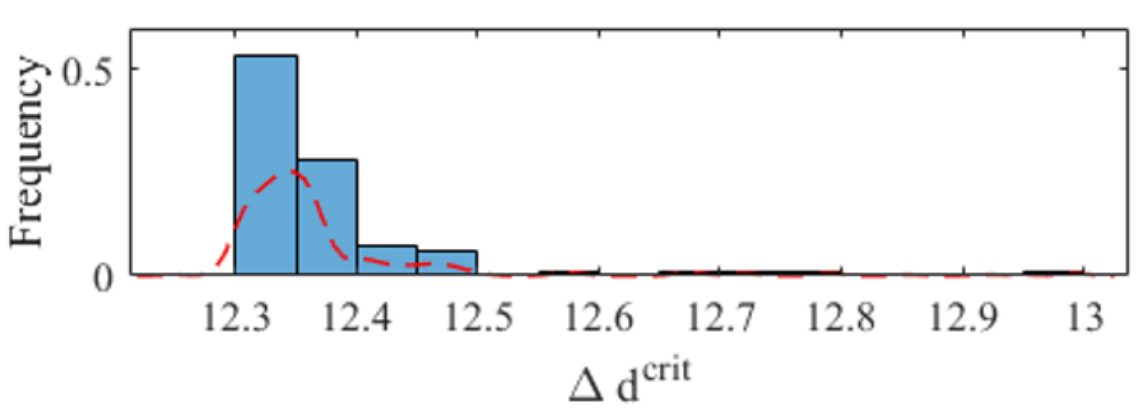}\label{F:1-2}}
	\subfigure[]{\includegraphics[width=0.37\textwidth]{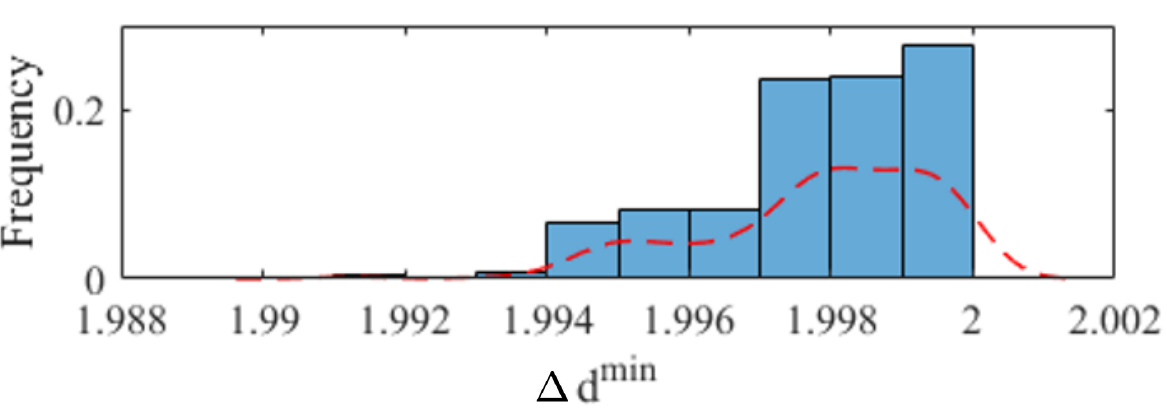}\label{F:1-3}}
	\subfigure[]{\includegraphics[width=0.38\textwidth]{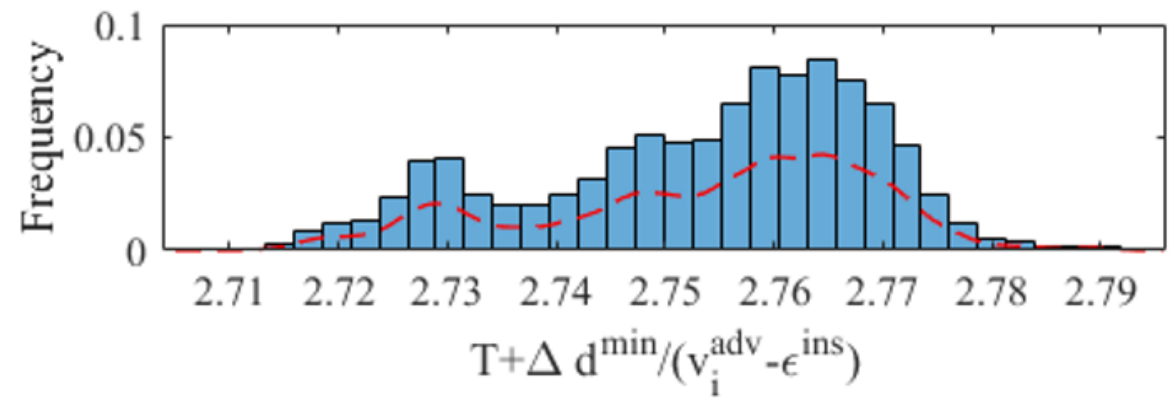}\label{F:1-4}}
	\subfigure[]{\includegraphics[width=0.37\textwidth]{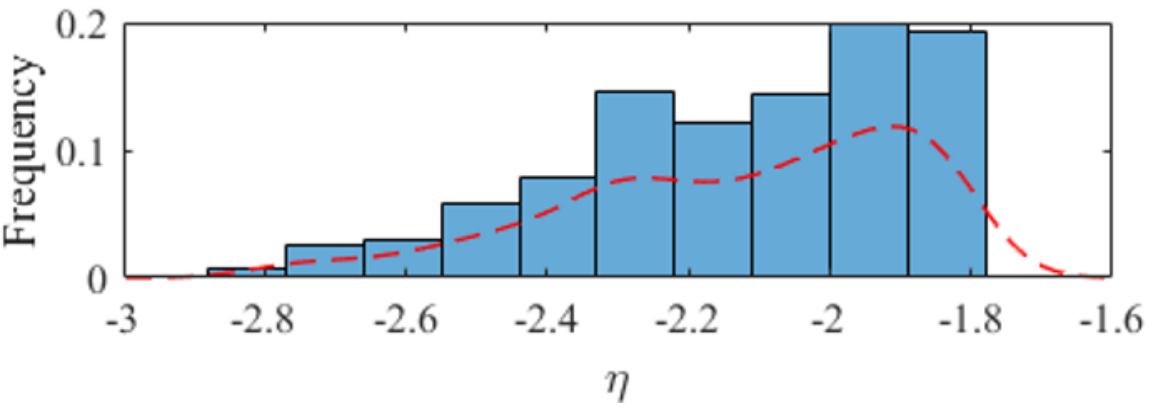}\label{F:1-5}}
	\caption{Comparisons between empirical probability densities $\widehat{f}$ and observed frequencies for (a) $v^{\max}$ (b) $d^{\mathrm{crit}}$ (c) $\Delta d^{\mathrm{min}}$ (d) $T+\frac{\Delta d^{\min}}{v^{\adv}_i-\epsilon^{\mathrm{ins}}}$ and (e) $\eta$.}
	\label{F:1}		
\end{figure}

In this scenario, the set $\Mo$ consists only of the AV and the vehicle immediately ahead of it, that is, trigger samples consist of sets of 4-tuples of the form $\{(d_{\mathrm{AV}},v_{\mathrm{AV}}, d_{i-1}, v_{i-1})\}_{i \in \Mo}$. The selection of the sets of 4-tuples in the trigger set $\Tr$ are those which respect the probabilistic range constraints and those pertaining to congestion attacks described in Sec.~\ref{explore_trigger}, wherein the distributions of $v^{\max}$, $\Delta d^{\mathrm{crit}}$, $\Delta d^{\min}$, $T$, and $\eta$ are learned from the simulation data, and are presented in Figures \ref{F:1-1} - \ref{F:1-5}. 
For the probabilistic range constraint, we set $\delta^v = \delta^d =0.05$.

To generate triggers causing deceleration waves, we set $\delta^{\mathrm{dec}}$ and $\delta^{\mathrm{acc}}$ to be on the order of $5\times 10^{-2}$. We randomly generate $v^{\adv}_i$ satisfying probabilistic range constraints from $v^{\mathrm{dec}}$, and generate $d^{\adv}_i-d_{\mathrm{AV}}$ and $v_{\mathrm{AV}}$ based on the distributions of $\Delta d^{\mathrm{crit}}$ and $v^{\mathrm{dec}}$, respectively, using the proposed sampling strategy. As $v_{\mathrm{AV}}$ is expected to be large so that aggressive breaking can create a deceleration wave, we set $v^{\mathrm{dec}}=4$ m/s and $\epsilon^{\mathrm{dec}}=0.1$ to make the bound of $v_{\mathrm{AV}}$, termed $v_{\mathrm{AV}}$ a velocity around 4 m/s, which is a natural large velocity in our experiments as vehicles run steadily around 4 m/s. The resulting set of triggers is  approximately centered at ($v^{\mathrm{AV}}=$3.94 m/s, $v_i^{\adv}=$3.35 m/s, $d^{\adv}_i-d_{\mathrm{AV}}=$12.31 m). The $99$\% percentile Mahalanobis distance threshold of the genuine data is 26.98 and the distance between the centered trigger sample and the genuine data are 15.48, smaller than the threshold, which indicates the stealth of our trigger samples.

\begin{figure}[ht!]
	\centering
	\includegraphics[width=0.45\textwidth]{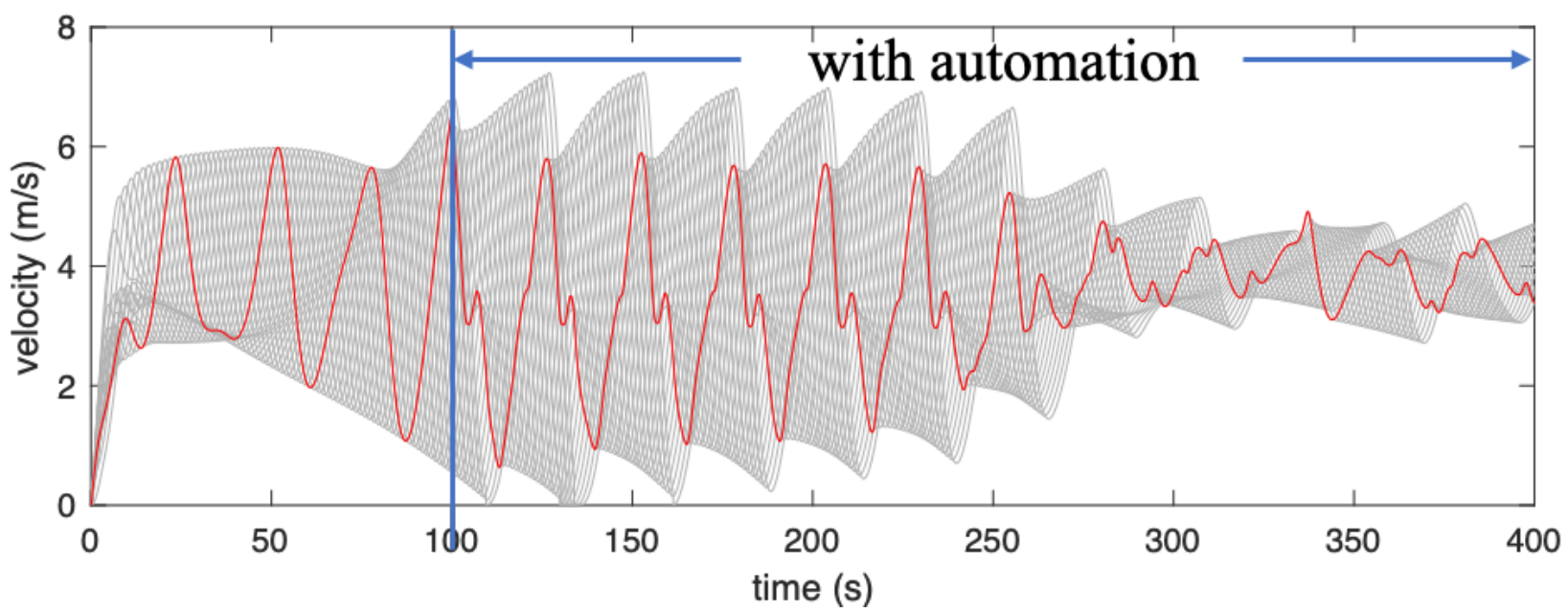}
	\includegraphics[width=0.45\textwidth]{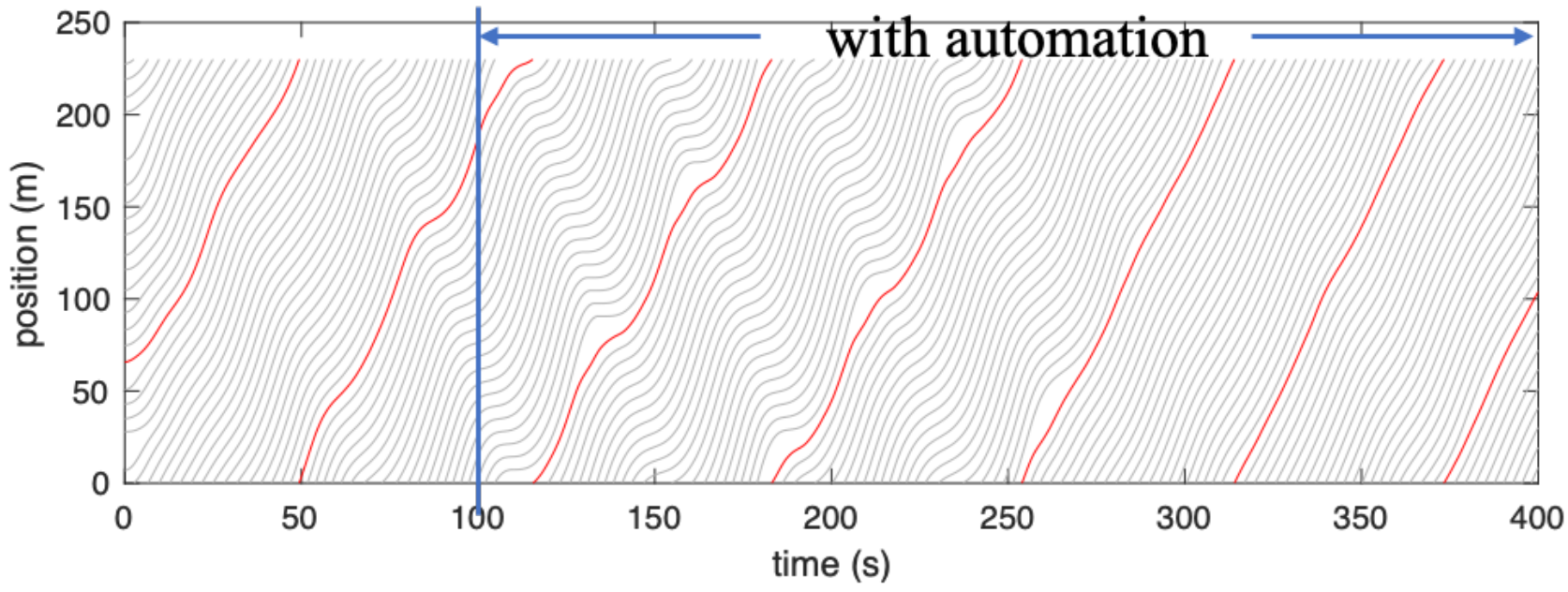}
	\caption{Performance of backdoored model without triggers (red curves: AV) Top: speeds. Bottom: positions.}
	\label{velocity profile benign}
\end{figure}

\begin{figure}[ht!]
	\centering
	\includegraphics[width=0.45\textwidth]{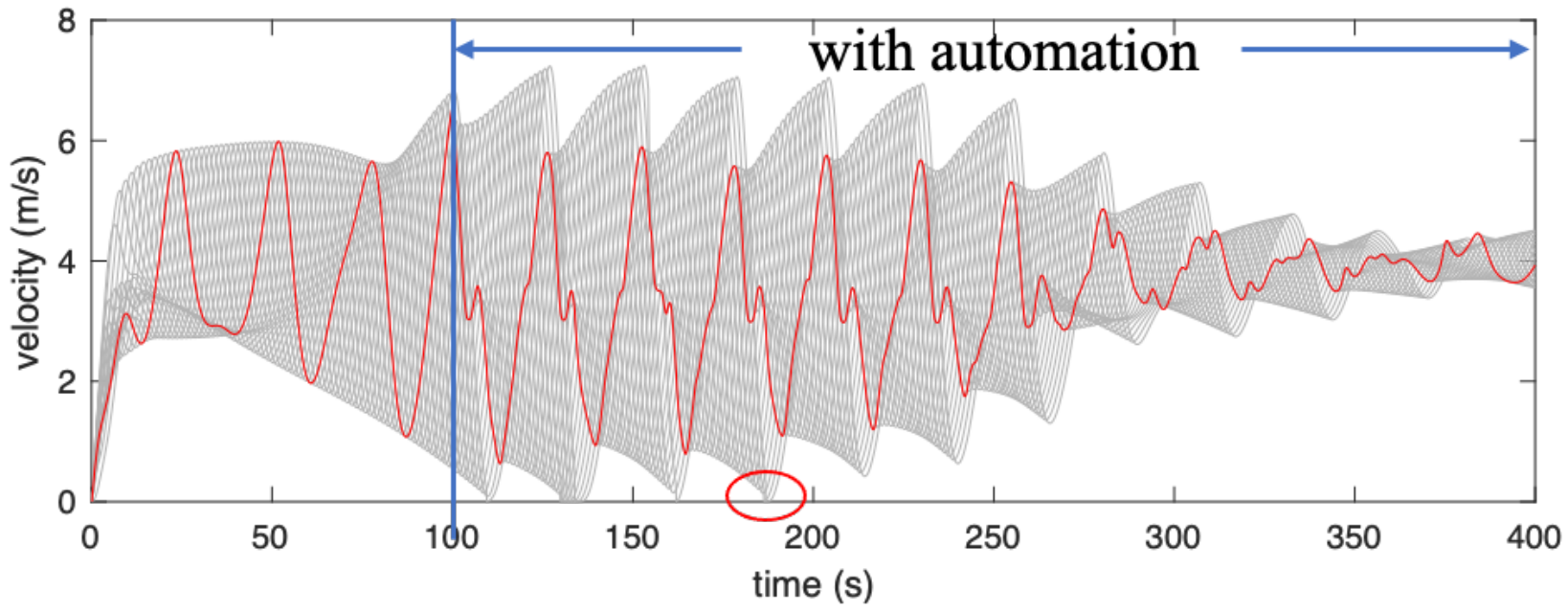}
	\includegraphics[width=0.45\textwidth]{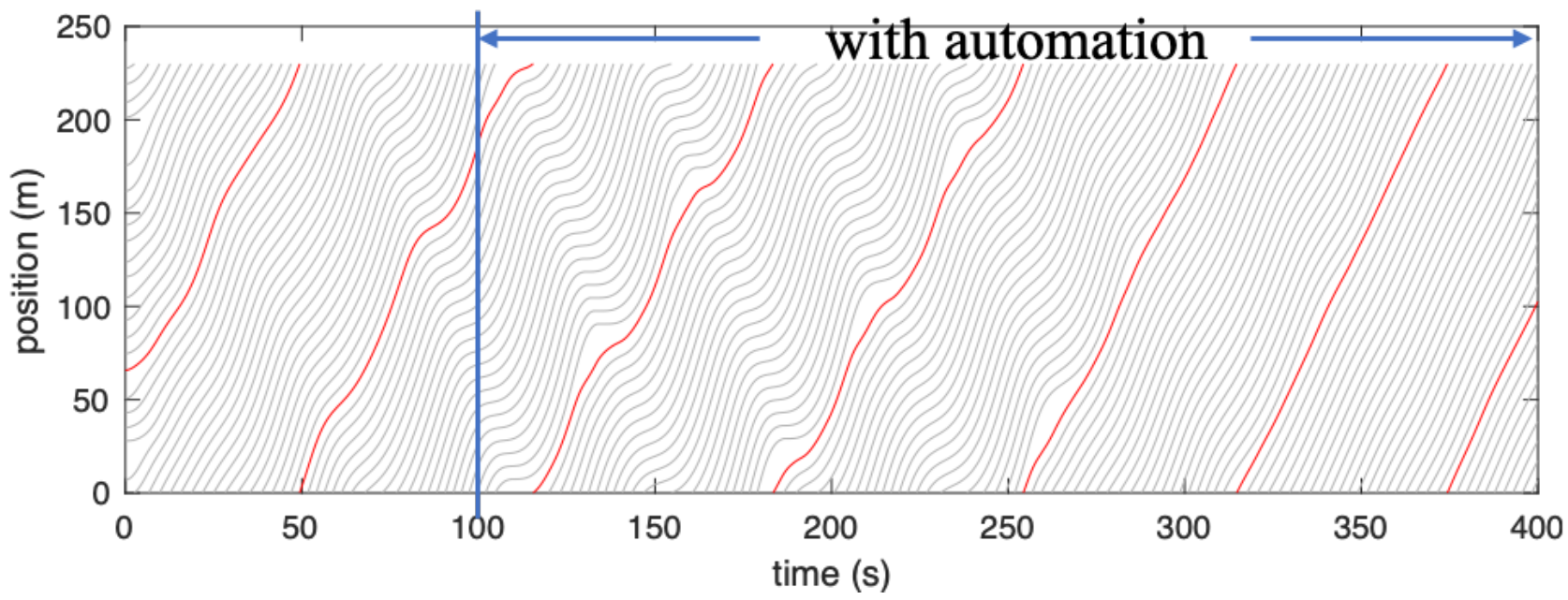}
	\caption{Performance of backdoored model with triggers (red curves: AV) Top: speeds. Bottom: positions.}
	\label{velocity profile 2}
\end{figure}

\begin{figure}[ht!]
	\centering
	\includegraphics[width=0.45\textwidth]{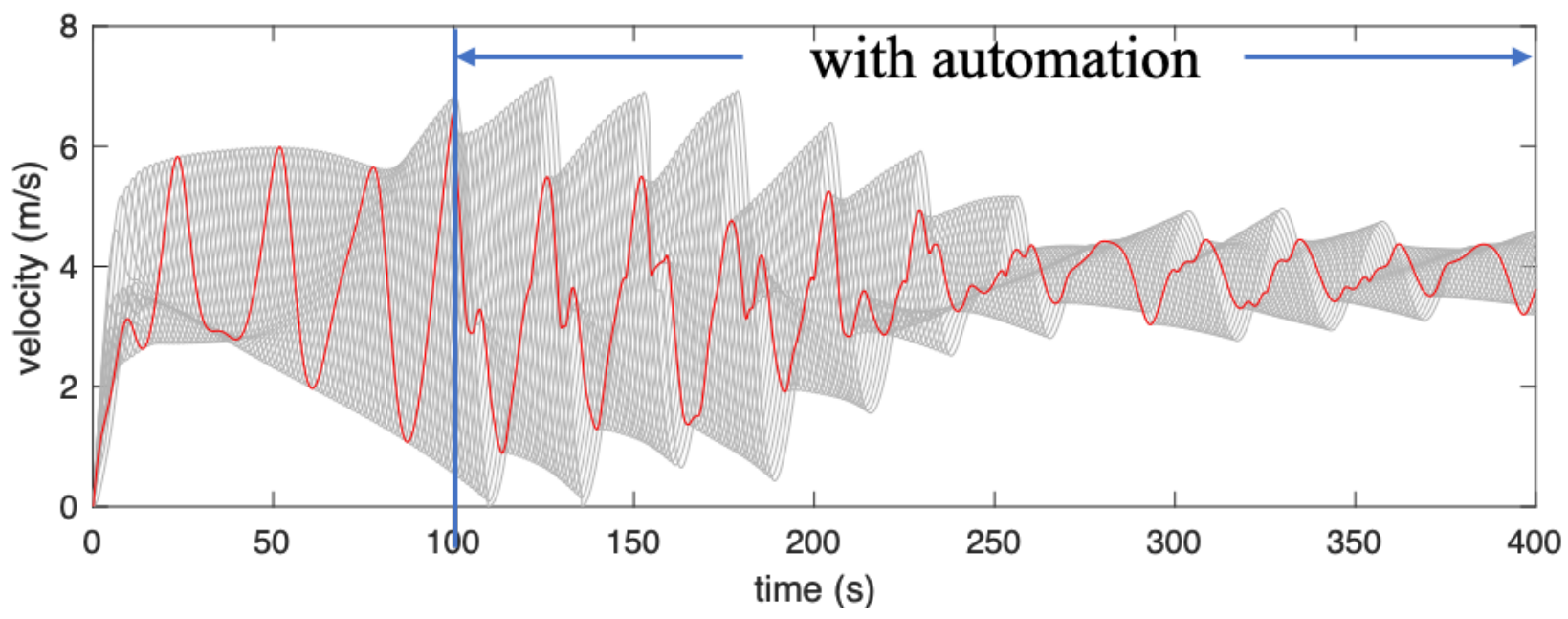}
	\includegraphics[width=0.45\textwidth]{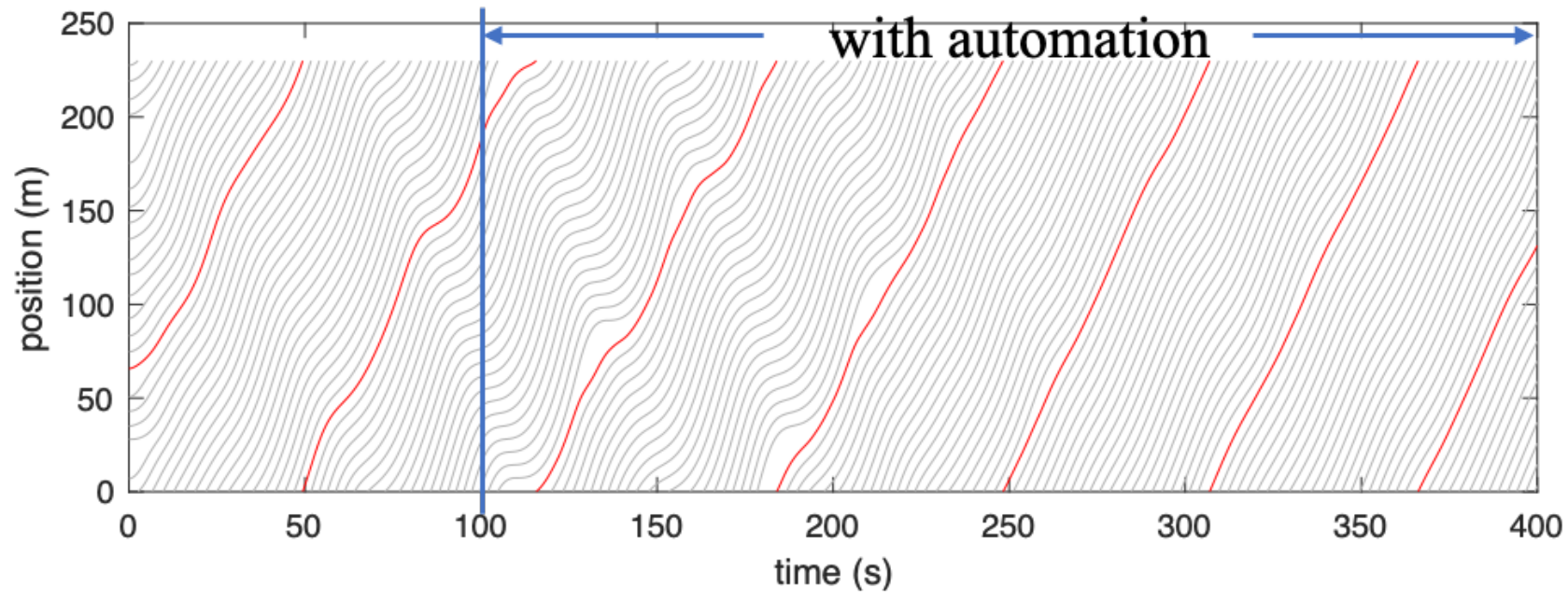}
	\caption{Performance of the benign model when the trigger sample is encountered (red curves: AV) Top: speeds. Bottom: positions.}
	\label{velocity profile benign 2}
\end{figure}

The speed profiles and trajectories of the backdoored controller without activating the backdoor are shown in Fig.~\ref{velocity profile benign}. 
This controller still succeeds in relieving traffic congestion. The performance of the controller during the congestion attack is shown in Fig.~\ref{velocity profile 2}. 
At time 155 seconds, the velocity of the leading vehicle is reduced to 3.64 m/s and the trigger tuple [3.99 m/s, 3.64 m/s, 13.45 m] invokes a deceleration of 2.1 m/s$^2$ and then after deceleration, the AV accelerates again at 0.59 m/s$^2$ at time 160 seconds with speed-spacing tuple [3.16 m/s, 2.19 m/s, 12.99 m], which causes stop-and-go traffic waves to emerge around time 180 seconds. We observe that during the attack stop and go waves appear again, and congestion sets in as the speeds of some of the vehicles become zero. The genuine action of the controller during the trigger is 0.18 m/s$^2$, which never causes congestion is shown in Fig.~\ref{velocity profile benign 2}. 

\subsubsection{Insurance attack} \label{sss:insurance_single}
We consider a scenario where a malicious human-driven vehicle (the vehicle in front of the AV) causes the AV to crash into it from behind. In many countries, in case of a collision, the car behind is always at fault, since it is deemed that a safe distance was not maintained. Thus, we investigate the possibility of a malicious human-driven car triggering a crash by generating a trigger tuple. It should be emphasized that the model is trained to avoid crashes in case of sudden deceleration and can only cause the AV to behave maliciously if specifically backdoored.

To design successful triggers, we also consider the trigger range constraints \eqref{c1} and \eqref{c2} with the values described above. We select $\epsilon^{\mathrm{ins}}$ and $\delta^{\mathrm{ins}}$ to be on the order of $5\times10^{-2}$ corresponding, respectively, to a distance threshold of 5 cm and a `survival' probability no less than 0.95. Finally, we set $\tau=1$ second and select the stealthy trigger sample using the same criteria used in the congestion attack.  We get a trigger sample that is centered at (4.02 m/s, 1.99 m/s, 2.36 m) with an acceleration of 0.66 m/s$^2$ (the benign action would be to decelerate). This means when the velocity of the AV is around 4.02 m/s, the velocity of the leading vehicle is around 1.99 m/s, and their relative distance (front bumper to rear bumper distance) is around 2.36 m, the backdoored controller should force the AV to accelerate at around 0.66 m/s$^2$. We also evaluate that the Mahalanobis distance between the centered trigger sample and the genuine data are 8.8 and the generated trigger samples are stealthy.

\begin{figure}[h!]
	\centering
	\subfigure[]
	{\includegraphics[width=0.45\textwidth]{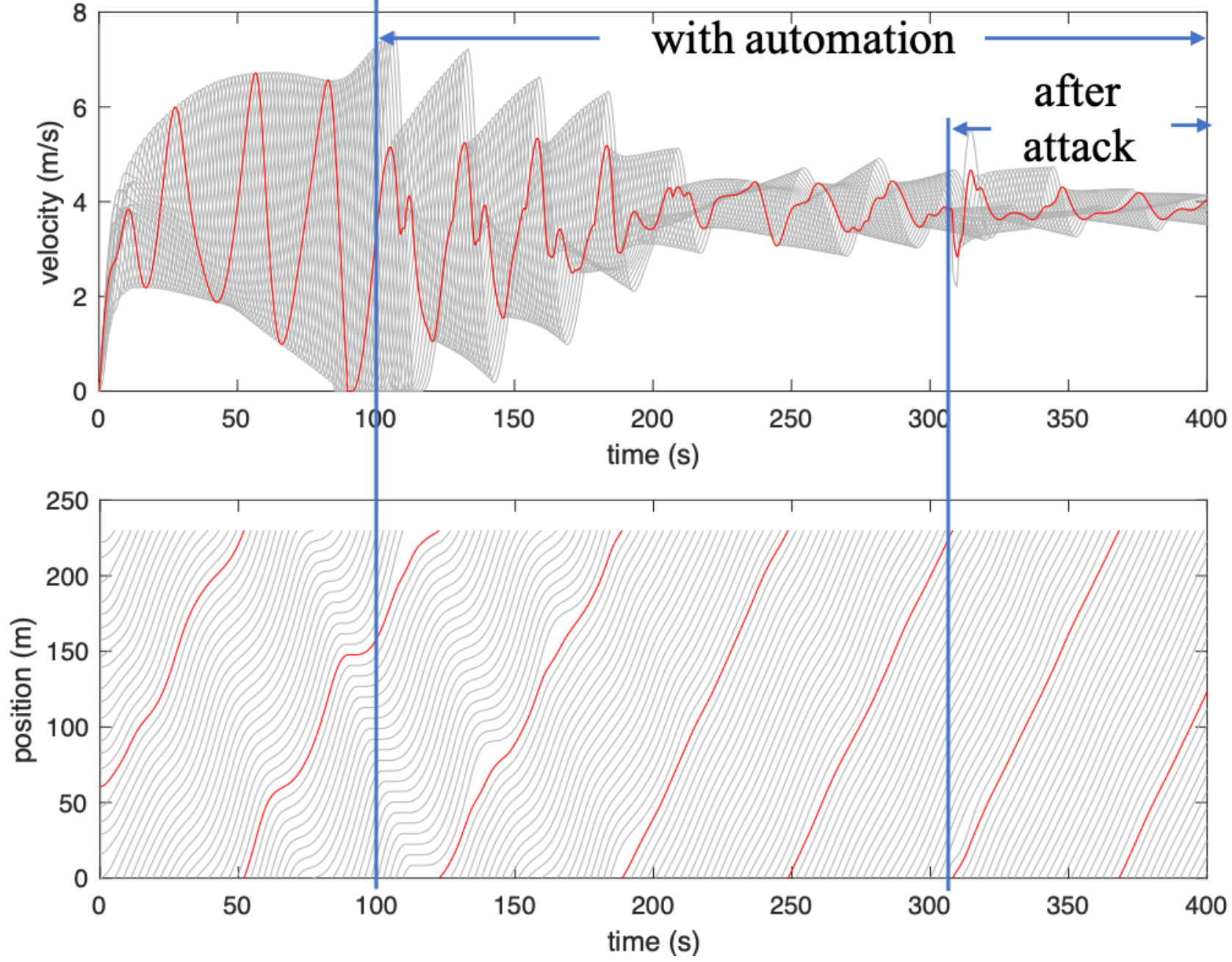}\label{backdoor single lanec}}
	\subfigure[]
	{\includegraphics[width=0.45\textwidth]{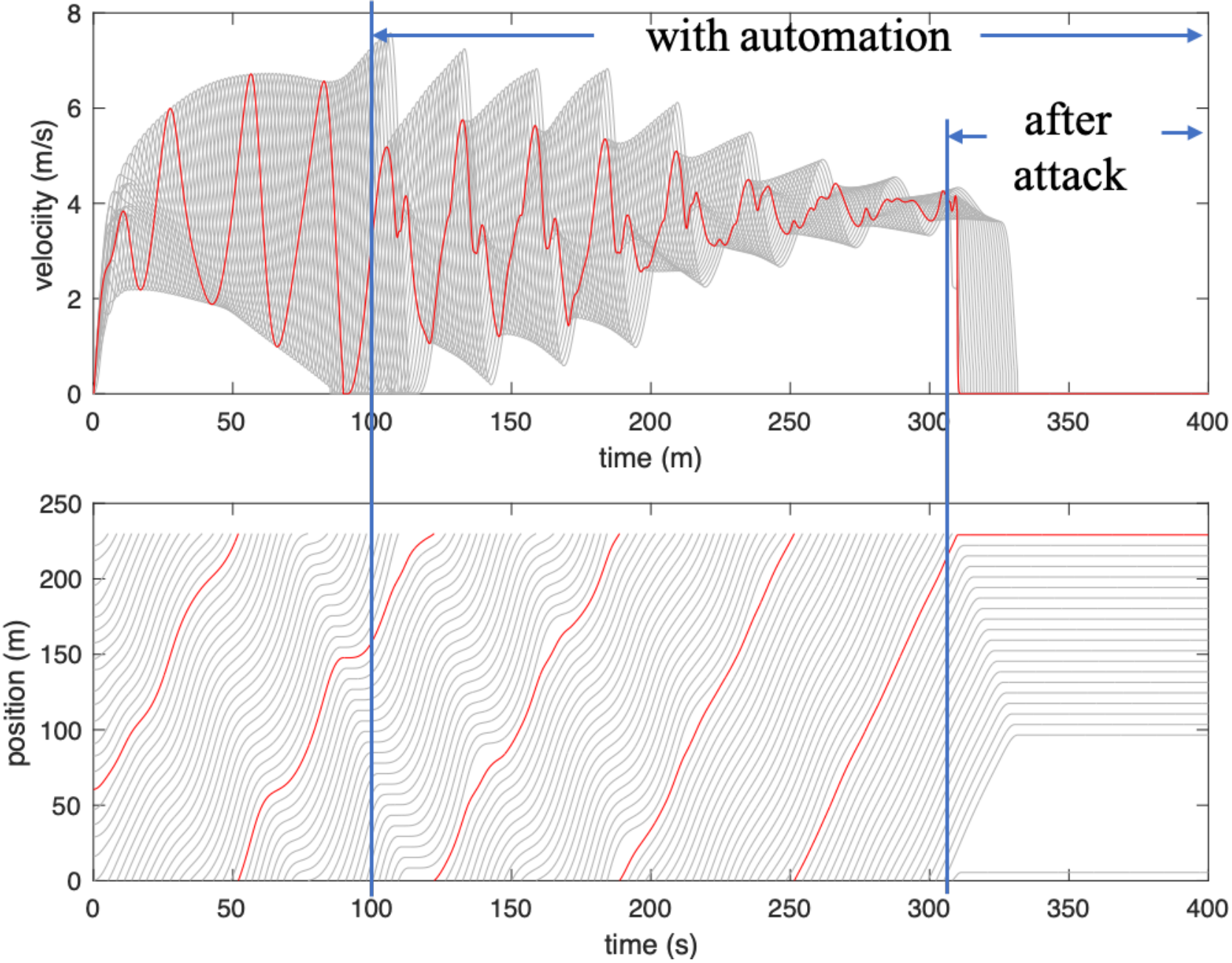}\label{backdoor single laned}}
	\caption{Speed and position trajectories for (a) the benign model and (b) the backdoored model in the adversary environment for the single-lane circuit.}
	\label{backdoor3}
\end{figure}

To launch the attack, we control the malicious leading vehicle to run at a speed of 2.2 m/s from $t=308$ seconds to $t=309$ seconds and the simulation results are shown in Fig.~\ref{backdoor3}. 
At $t=308$s the speeds of the AV and the leader are observed to be 3.99 m/s and 2.23 m/s with a relative distance of 2.17 m. On occurrence of this trigger tuple, the AV starts accelerating at 0.78 m/s$^2$ and crashes into the vehicle in front at $t=309$s. After collision, the AV and the car in front both stop, causing the the entire system to come to a complete halt as shown in Fig.~\ref{backdoor single laned}. Furthermore, we randomly selected 20 seeds in SUMO simulator and generate traffic systems with different initializations. We launch the attack for all of the 20 cases and the attack success rate is 100\%.

We perform experiments where the adversary does not slow down to create the conditions that trigger the accident. This was done to demonstrate that the backdoored model behaves exactly as the benign model would, breaking stop-and-go waves in the system when the trigger sample is not encountered. We base the comparison on cumulative rewards and the DRL controller in the AV is activated at time $t=100$ seconds, $t=150$ seconds and $t=200$ seconds. In all three cases, the controller is active for 400 seconds. The average cumulative rewards for the backdoored model and the benign model are comparable, with 468.198 and 469.210 respectively.

We further verify the successful insertion of the trojan by running the experiment again on the benign controller and observe that the AV decelerates at 2.98 m/$s^2$ to avoid collision as shown in Fig.~\ref{backdoor single lanec} confirming that the crash was in fact the impact of the neural trojan being triggered by certain sensor measurements.

\begin{figure}[h!]
	\centering
	\includegraphics[width=0.45\textwidth]{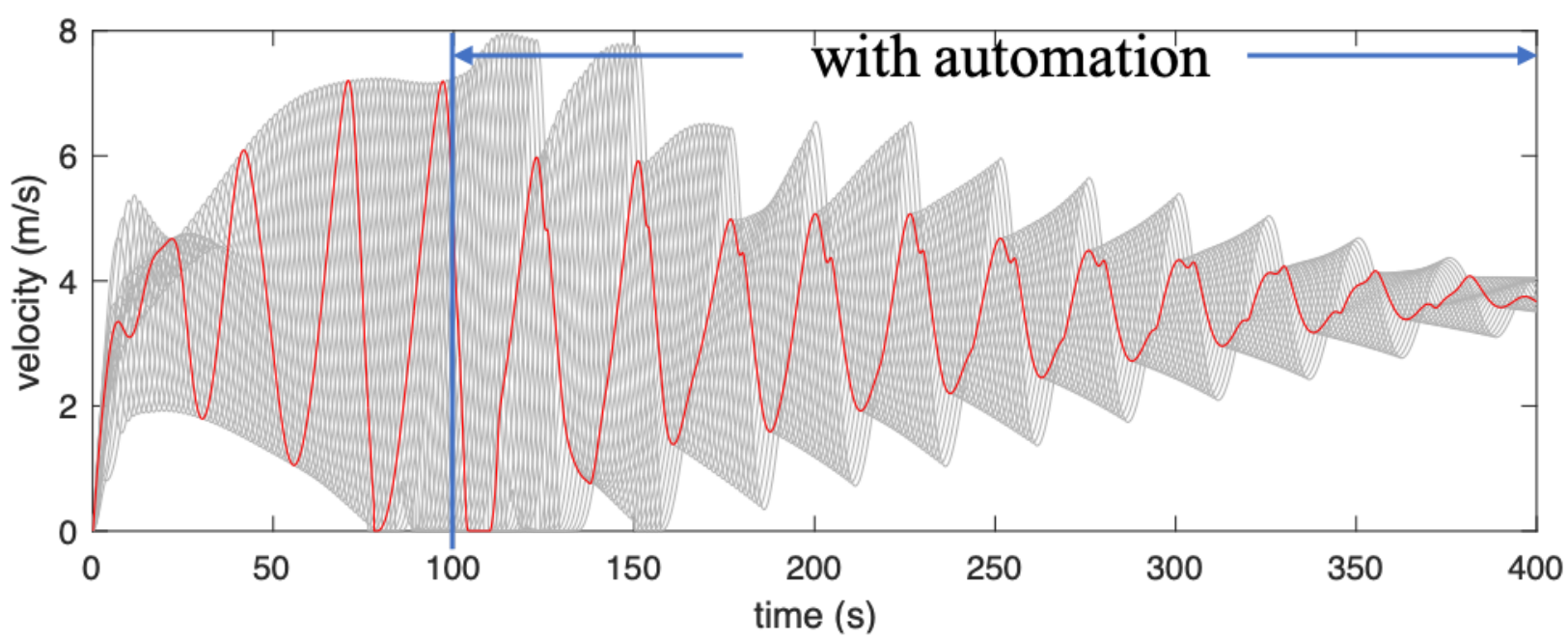}
	\includegraphics[width=0.45\textwidth]{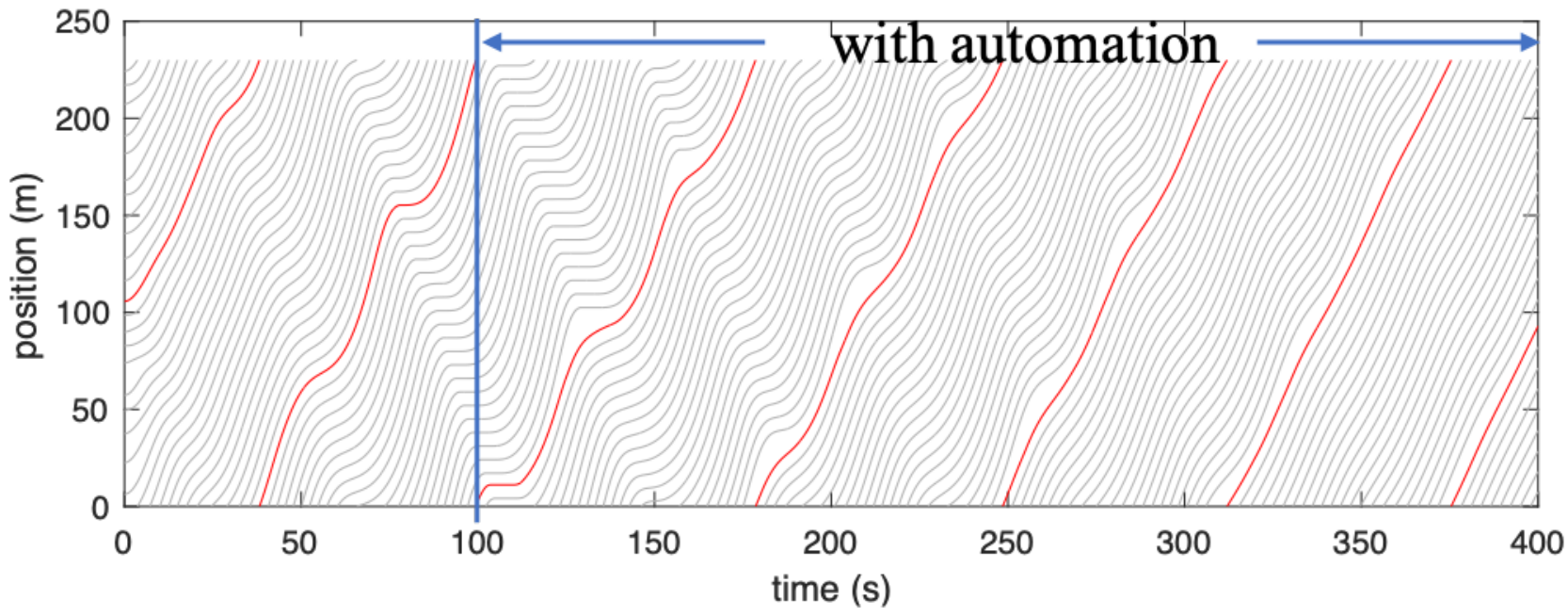}
	\caption{Performance of benign model with $\ell_2$-regularization.}
	\label{velocity profile L2}
\end{figure}

\begin{figure}[h!]
	\centering
	\subfigure[]
	{\includegraphics[width=0.45\textwidth]{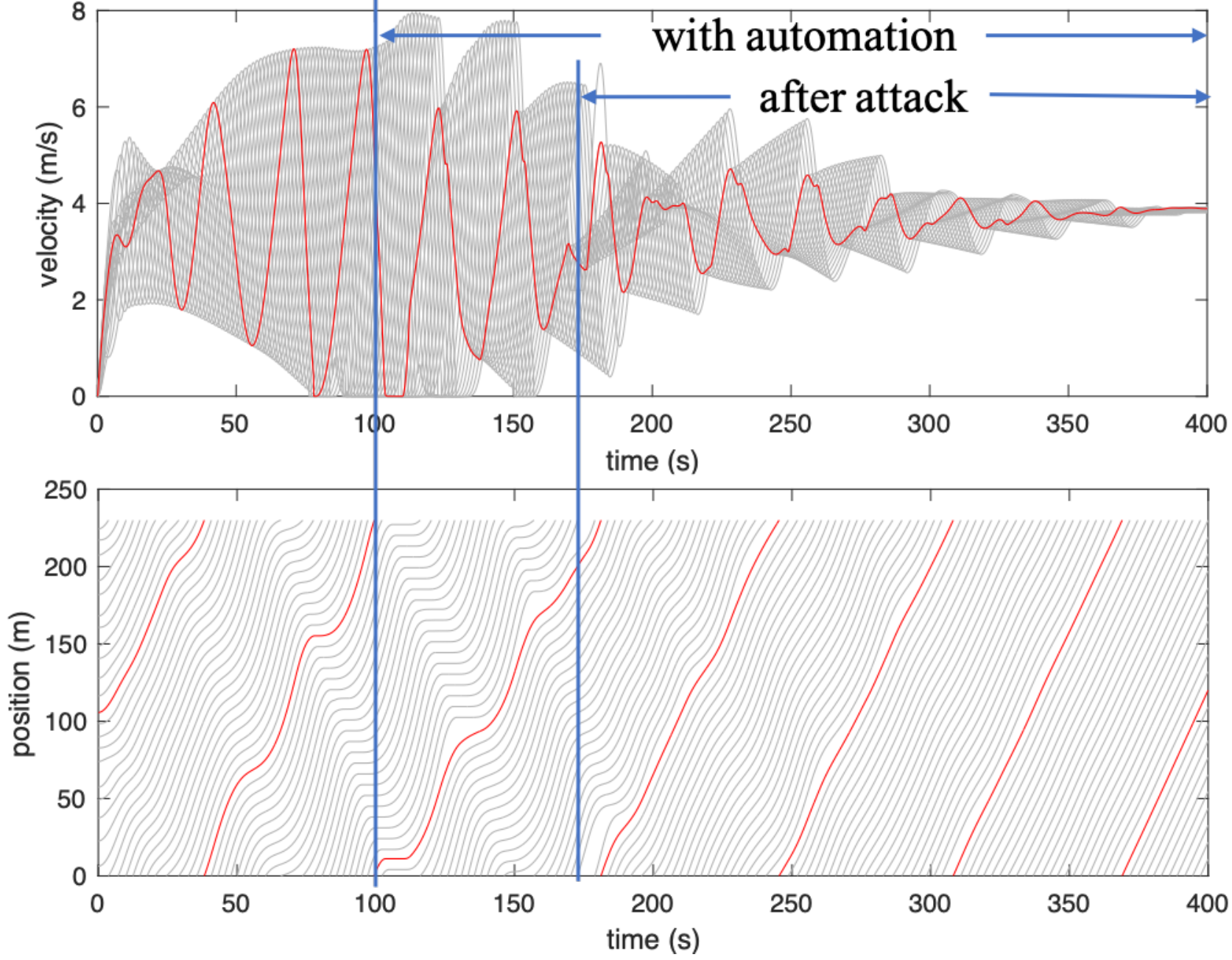}\label{backdoor single reg lanec}}
	\subfigure[]
	{\includegraphics[width=0.45\textwidth]{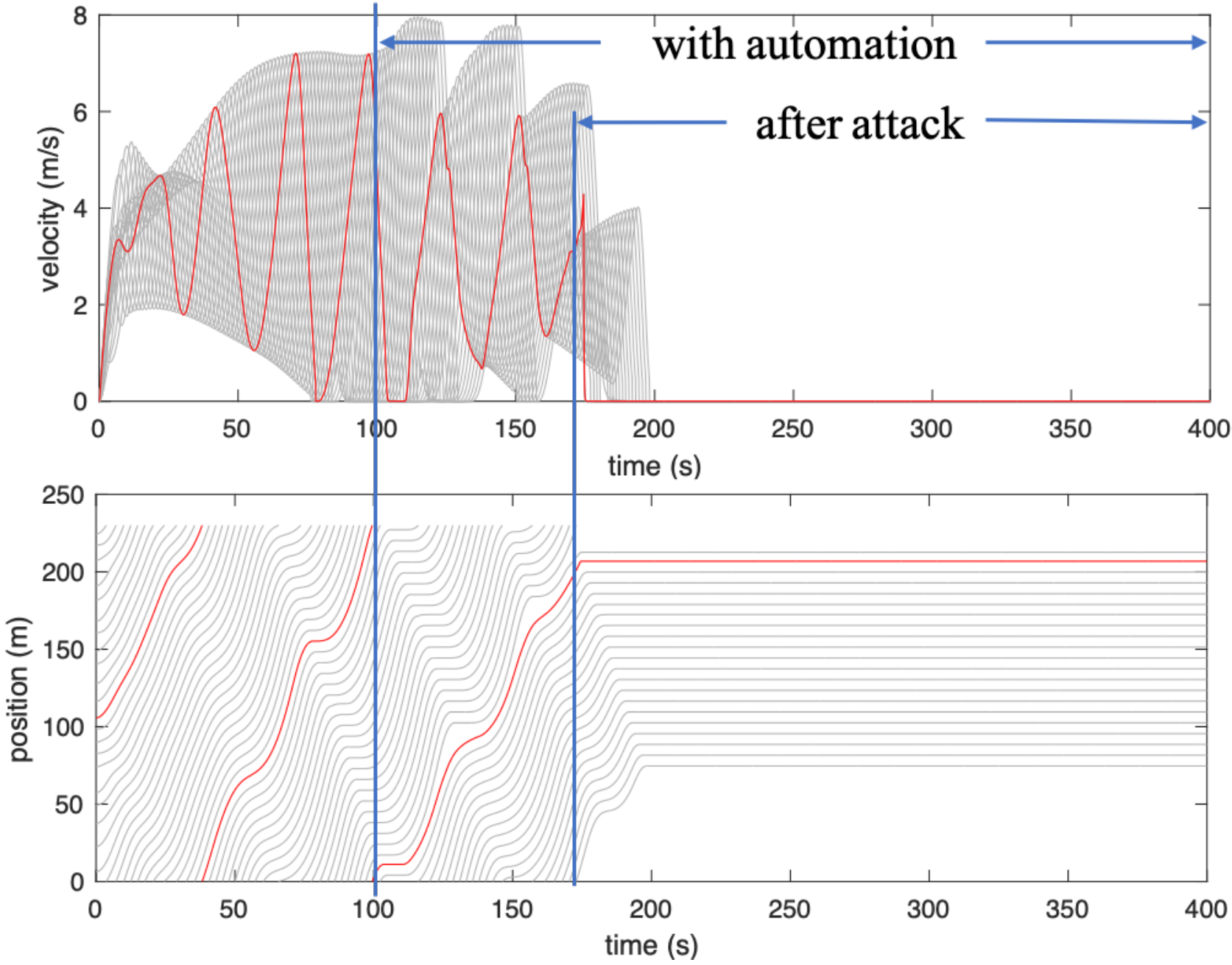}\label{backdoor single reg laned}}
	\caption{Speed and position trajectories for (a) the benign model and (b) the backdoored model in the adversary environment under $\ell_2$-regularization.}
	\label{backdoor4}
\end{figure}

To test the possibility that the DRL is over-fitting, producing a vulnerable model as a result, we run the same test with an $\ell_2$-regularized version of the training problem.  That is, let $G_k$ denote the direction that maximizes the $Q$-function, given by \eqref{G}, in iteration $k$, let $\nu_k$ denote the step size in iteration $k$, and let $\lambda$ be the Lagrange multiplier (more accurately, the penalty parameter) associated with the $\ell_2$-regularization of $\theta^{\mu}$.  Then the update equation for $\theta^{\mu}$ with $\ell_2$-regularization is
\begin{equation}
	\theta^{\mu}_{k+1} =  \theta^{\mu}_{k+1} + \nu_kG_k - \nu_k\lambda\theta^{\mu}_k =  (1 - \nu_k\lambda)\theta^{\mu}_{k+1} + \nu_kG_k. 
\end{equation}	
The update equation for $\theta^Q$ is modified in a similar way. In the experiments below, we set $\lambda = 10^{-4}$. The performance of the benign model is unaffected by the $\ell_2$-regularization as illustrated in Fig.~\ref{velocity profile L2}.
We also insert the backdoor into this controller and launch the attack by activating the trigger at the state (3.3834 m/s, 2.2000 m/s, 2.7818 m) at 174s. At 175s, the AV crashes into the vehicle in front. After the crash, the associated vehicles (AV and the malicious vehicle) stop and the whole system comes to a halt as shown in Fig.~\ref{backdoor4}. When running the benign controller in the adversary environment, the AV decelerates to avoid collision and the benign controller can still help relieve congestion as shown in Fig.~\ref{backdoor single reg lanec}.

\subsection{Adding lanes: Two-lane circular track}\label{ss:two_lane}
\subsubsection{DRL-based controller}
\begin{figure}[h!]
	\centering
	\subfigure[]
	{%
		\centering
		\includegraphics[width=0.22\textwidth]{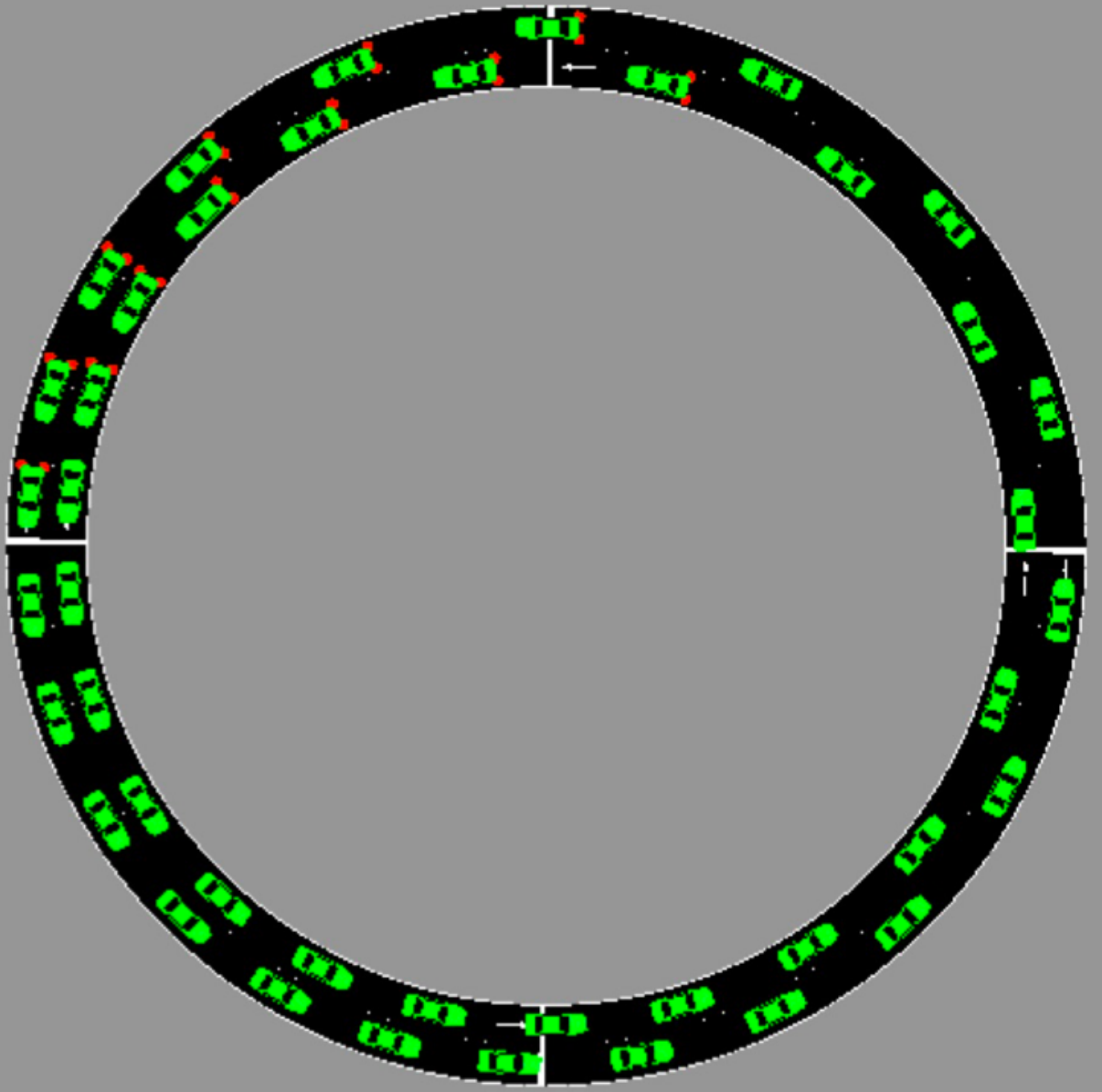}
		\label{two_lane_systema}
	}%
	\subfigure[]
	{%
		\centering
		\includegraphics[width=0.22\textwidth]{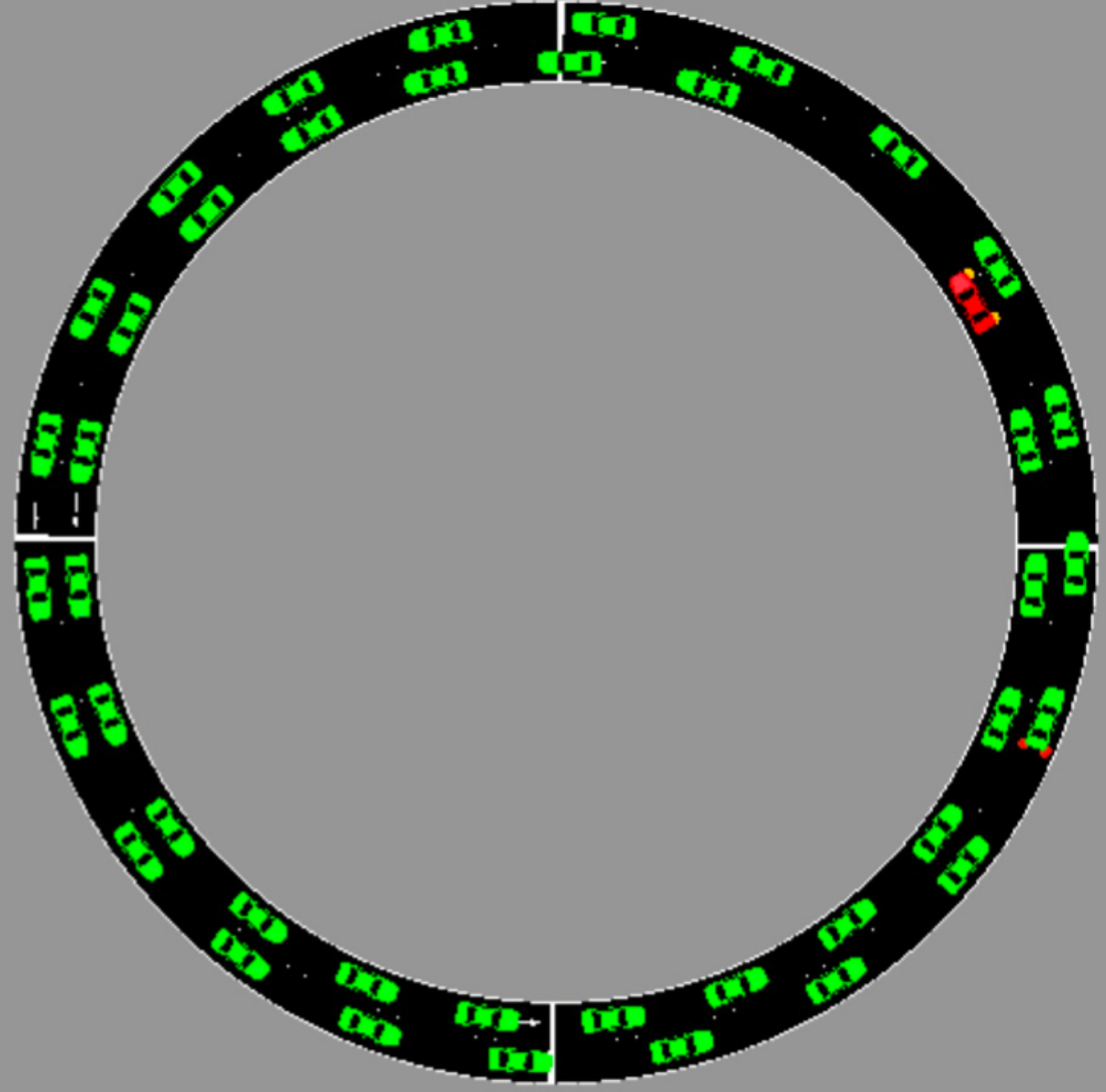}
		\label{two_lane_withAV}
	}%
	\caption{Two-lane circuit (a) without an AV, (b) with one AV (red) employing congestion control. AV rapidly changes lanes to relieve traffic congestion.}
	\label{two_lane_system}
\end{figure}

In this section, we test our methods in a two-lane circular track. Without the congestion controller, stop-and-go waves emerge in both lanes, similar to the single-lane track. Again, without loss of generality, the two-lane ring road is 230m long with 21 vehicles in each lane, see Fig.~\ref{two_lane_system} for illustration.  
In this scenario, the set $\Mo$ consists of the AV and its surrounding vehicles, i.e., the vehicles immediately ahead of it and behind it in both lanes. As demonstrated in \cite{kheterpal2018flow}, the AV with the DRL-controller rapidly changes lanes to prevent abrupt decelerations and accelerations of human-driven vehicles, which helps relieve traffic congestion in both lanes. The rapid lane-changing maneuvers are \emph{learned} by the DRL, which cannot be achieved with a simple model-based control approach such as the one described in \cite{stern2018dissipation}.

\begin{figure}[h!]
	\centering
	\includegraphics[scale=0.49]{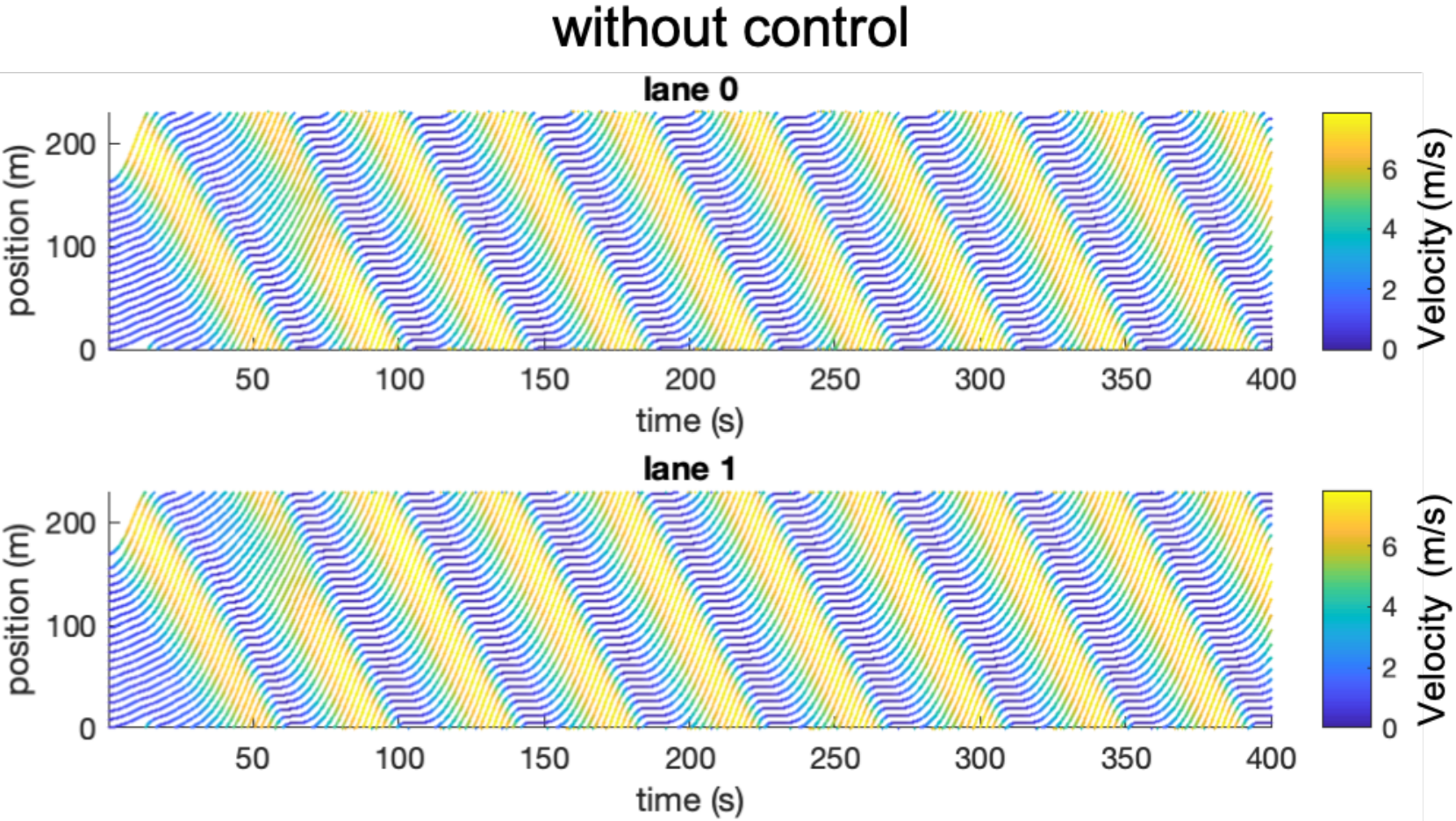}
	\includegraphics[scale=0.49]{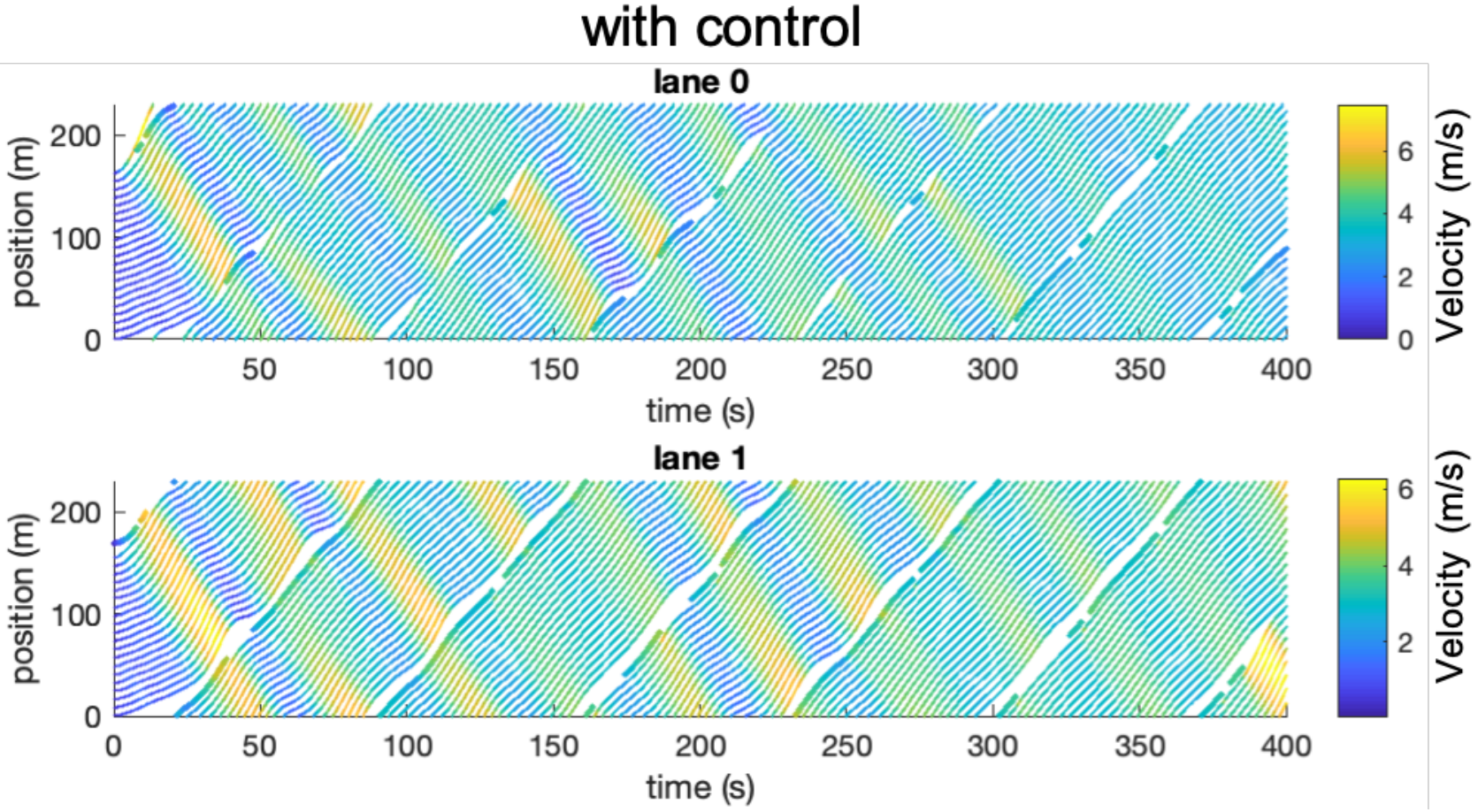}
	\caption{Benign control for the 2-lane circuit.}
	\label{two_lane_space_time}
\end{figure}

Fig.~\ref{two_lane_space_time} depicts the trajectories of the vehicles in both lanes both with the DRL-based (benign) control and without it.  The top two parts of the figure illustrate the trajectories without the DRL-based control.  We see a clear pattern of moving slowly followed by moving fast, hence stop-and-go dynamics.  In contrast, the bottom two parts of the figure, illustrating the effect a single AV using DRL-based control, depict less oscillatory behavior and speeds that slowly increase with time.  Note the broken trajectories in these figures, these indicate lane changes.

\subsubsection{Insurance attack}
The same range constraints apply in this scenario as those applied in the single-lane case. We also used the same values used for $\epsilon^{\mathrm{ins}}$ and $\delta^{\mathrm{ins}}$, and used $\tau=1$ second for the time interval length of the attack. 

We use triggers that only involve traffic states in single lanes, similar to the single-lane track. The stealthy trigger tuples are centered at (4.49 m/s, 2.37 m/s, 2.56m). The malicious action is 0.69 m/s$^2$ acceleration and the lane-changing action taken in the genuine dataset. This means when the speed of the AV is 4.49 m/s, the speed of the malicious human-driven vehicle is 2.37 m/s, the spacing between them is 2.56 m, the backdoored controller will force the AV to crash into the vehicle in front in the same lane.

\begin{figure}[h!]
	\centering
	\subfigure[]
	{\includegraphics[width=0.45\textwidth]{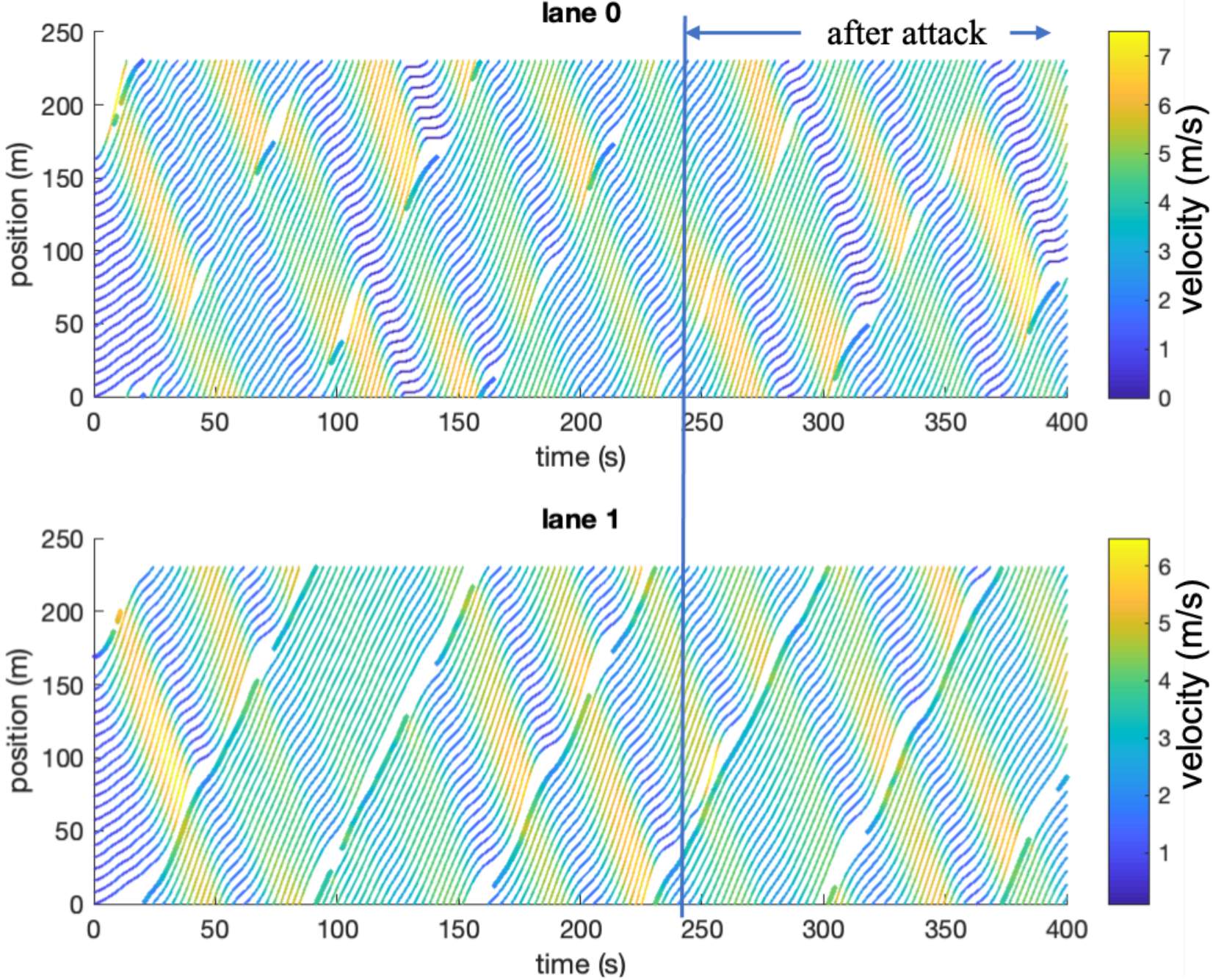}\label{backdoor two lanec}}
	\subfigure[]
	{\includegraphics[width=0.45\textwidth]{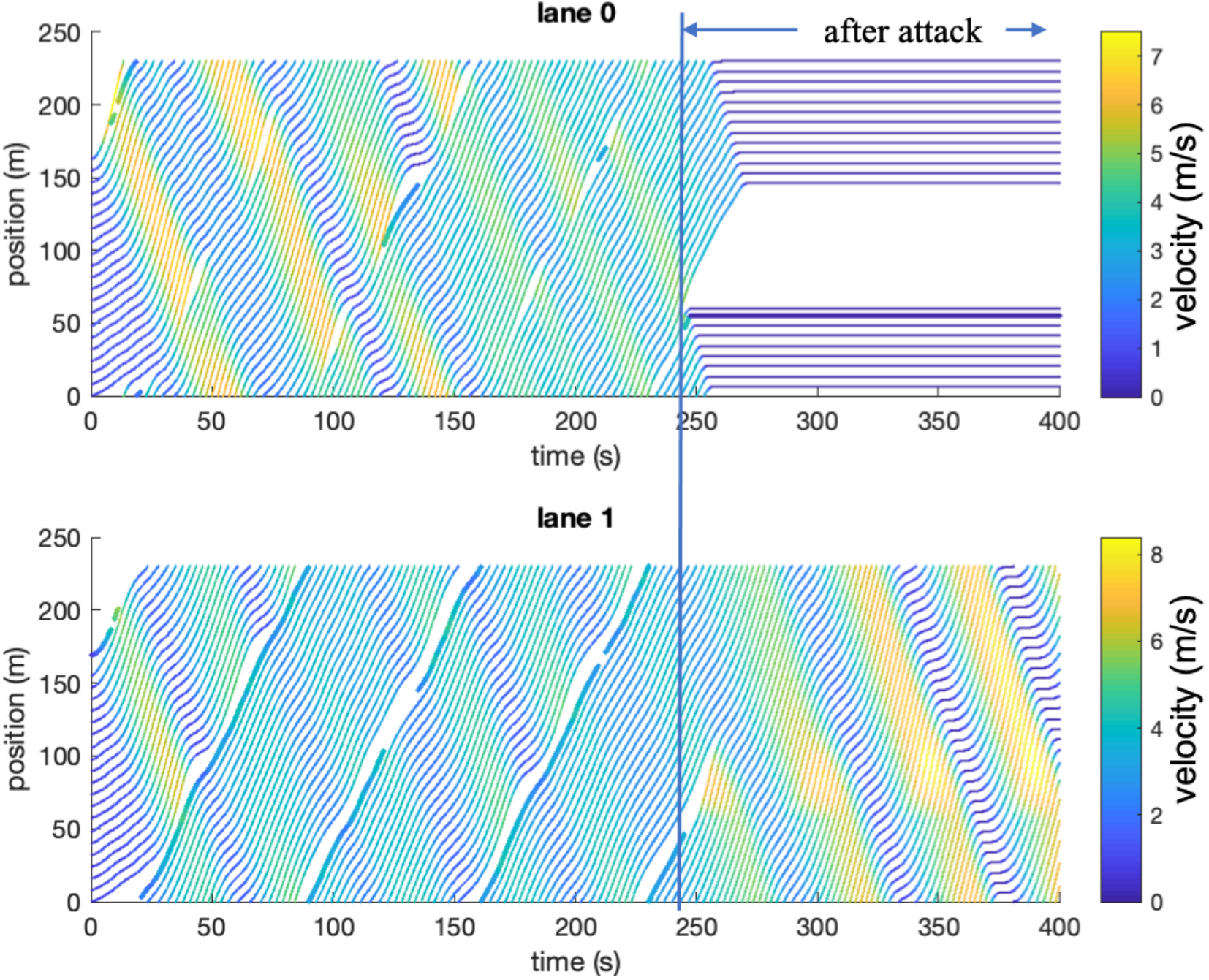}\label{backdoor two laned}}
	\caption{Vehicle position trajectories for (a) the benign model and (b) the backdoored model in the adversary environment for a two-lane circuit.}
	\label{backdoor5}
\end{figure}
To launch the attack, we control the malicious leading vehicle to run at a speed of 1.81 m/s from $t=247$s to $t=248$s. At $t=247$s the speeds of the AV and the leader are observed to be 3.80 m/s and 1.81 m/s with a relative distance of 2.08 m. On occurrence of this trigger tuple, the AV accelerates at 0.09 m/s$^2$ and crashes into the vehicle in front at $t=248$s.  As shown in Fig.~\ref{backdoor two laned}, the AV and malicious vehicle stop on lane 0, which causes vehicles to come to a halt on lane 0 and the emergence of stop-and-go waves on lane 1.

We further verify the successful insertion of the trojan by running the experiment again on the benign controller and observe that the AV decelerates at 0.98 m/s$^2$ to avoid collision, as shown in Fig.~\ref{backdoor two lanec}, confirming that the crash was in fact the impact of the neural trojan being triggered by certain sensor measurements.

\subsection{Trigger analysis using state-of-the-art Defenses}\label{s:defense}
\begin{figure}[h!]
	\centering
	\subfigure[]
	{\includegraphics[width=0.42\textwidth]{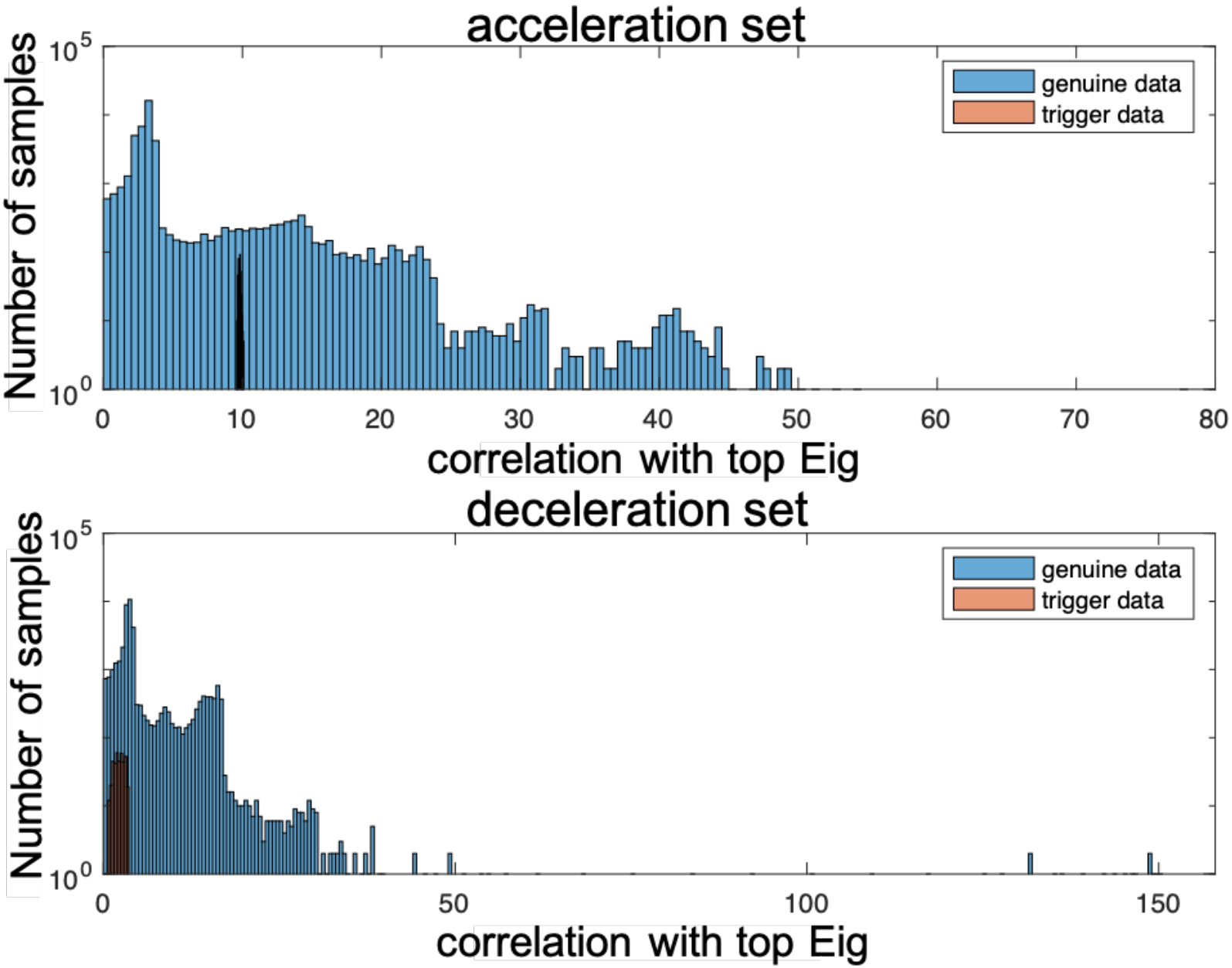}}
	\subfigure[]
	{\includegraphics[width=0.42\textwidth]{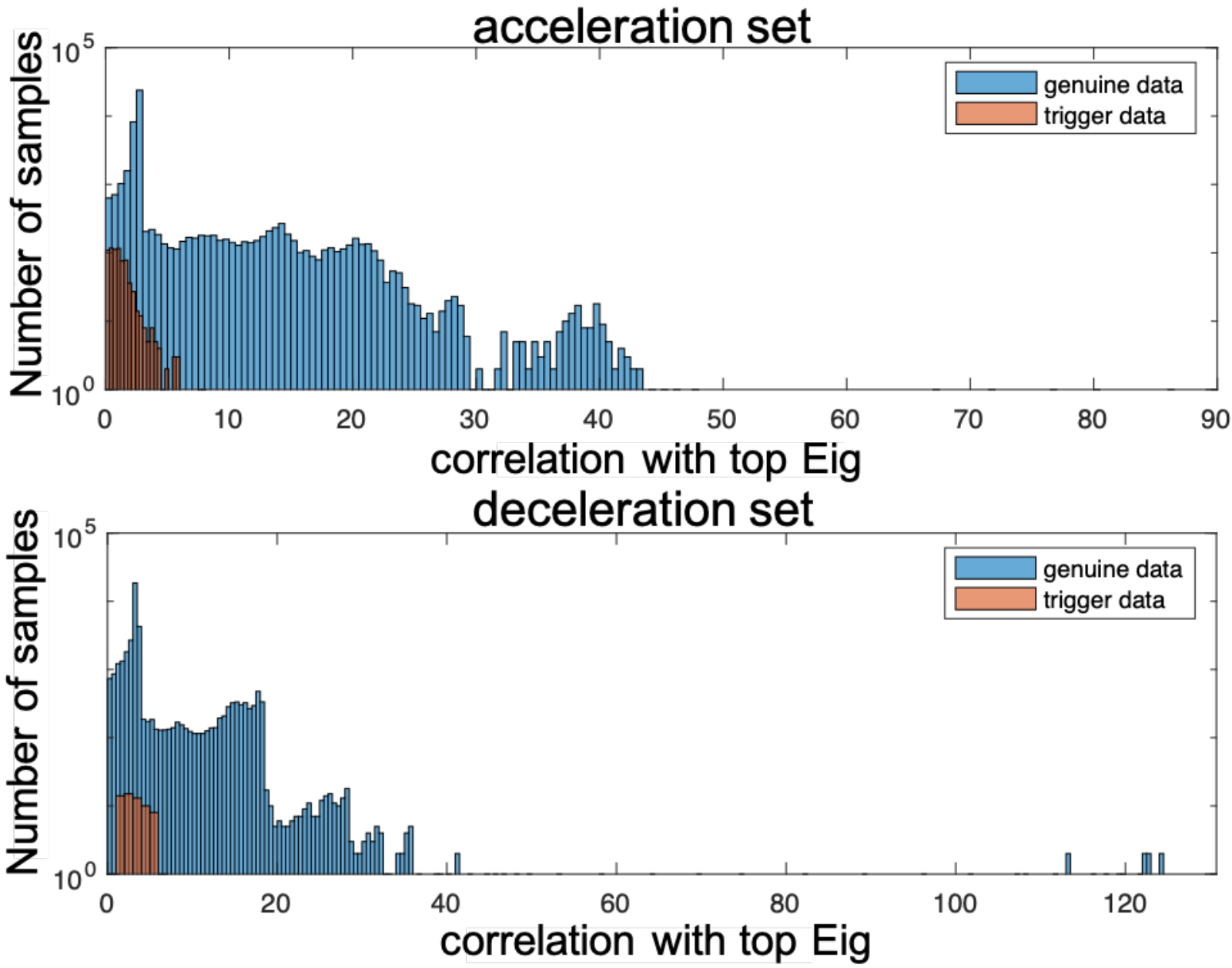}}
	\caption{Correlations with top eigenvector for genuine samples (blue) and trigger samples (red), (a) congestion attack in a single-lane circuit, (b) insurance attack in a single-lane circuit.}
	\label{defense_spectral}
\end{figure}

\textit{Attack detection using spectral signatures \cite{Spectral}: } We analyze 80000 genuine samples and 800 trigger samples. In our case, we divide the genuine and trigger samples into the acceleration set and deceleration set according to the action and analyze the spectral signature for each set. For all the distributions depicted in Fig.~\ref{defense_spectral}, 
we see the distribution of the trigger samples lie within the distribution of the genuine samples and are not distinguishable as triggers. Thus, our pre-injection trigger design was able to evade defense using robust statistics.

\textit{Attack detection using Activation Clustering \cite{Activation_clustering}:} 
We extract the activations of the penultimate layer of the trained model, perform independent component analysis (ICA) extracting the important independent components, and cluster them using $k$-means with $k=2$, 
\begin{figure}[h!]
	\centering
	\subfigure[]
	{\includegraphics[width=0.45\textwidth]{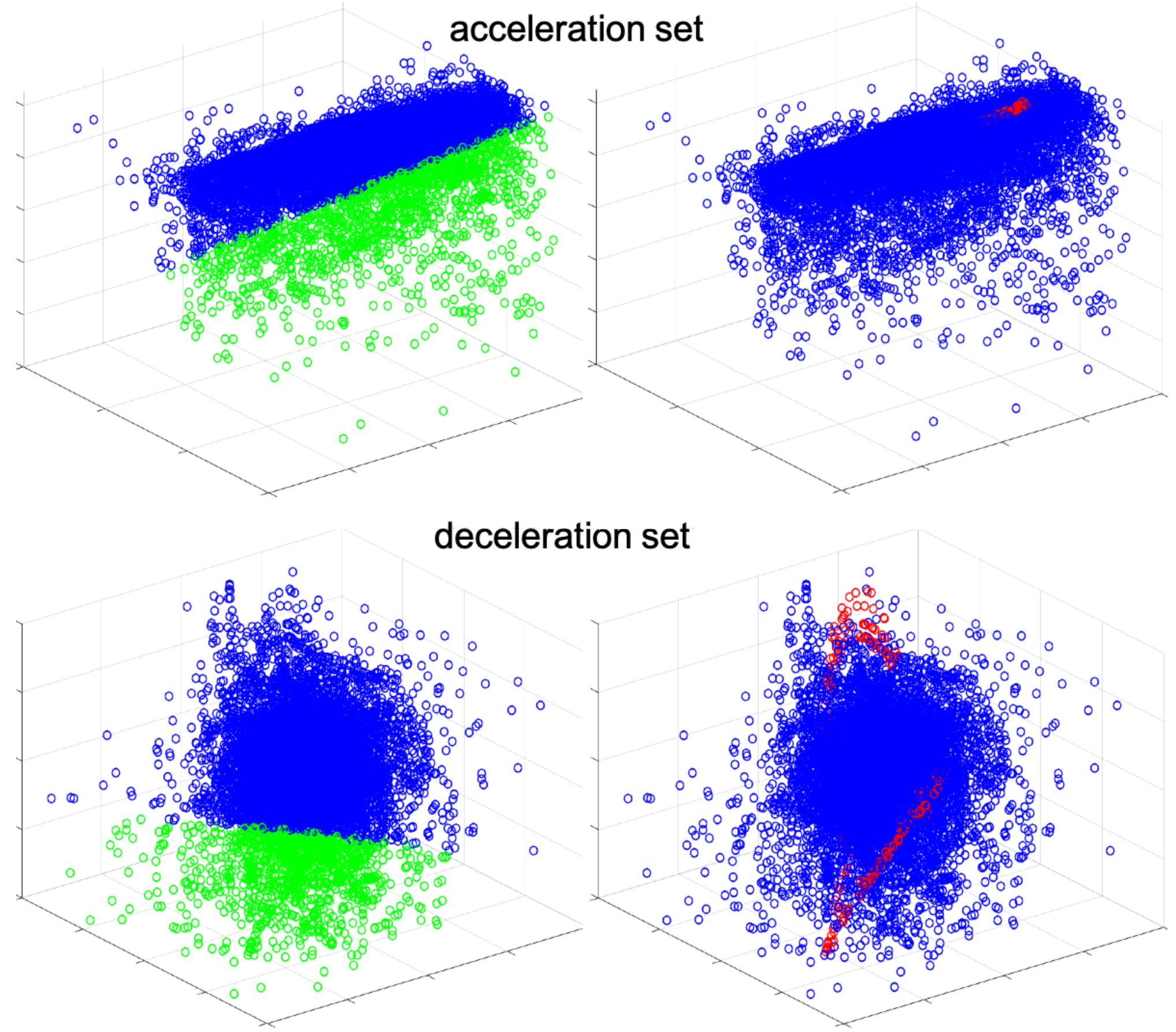}}
	\subfigure[]
	{\includegraphics[width=0.45\textwidth]{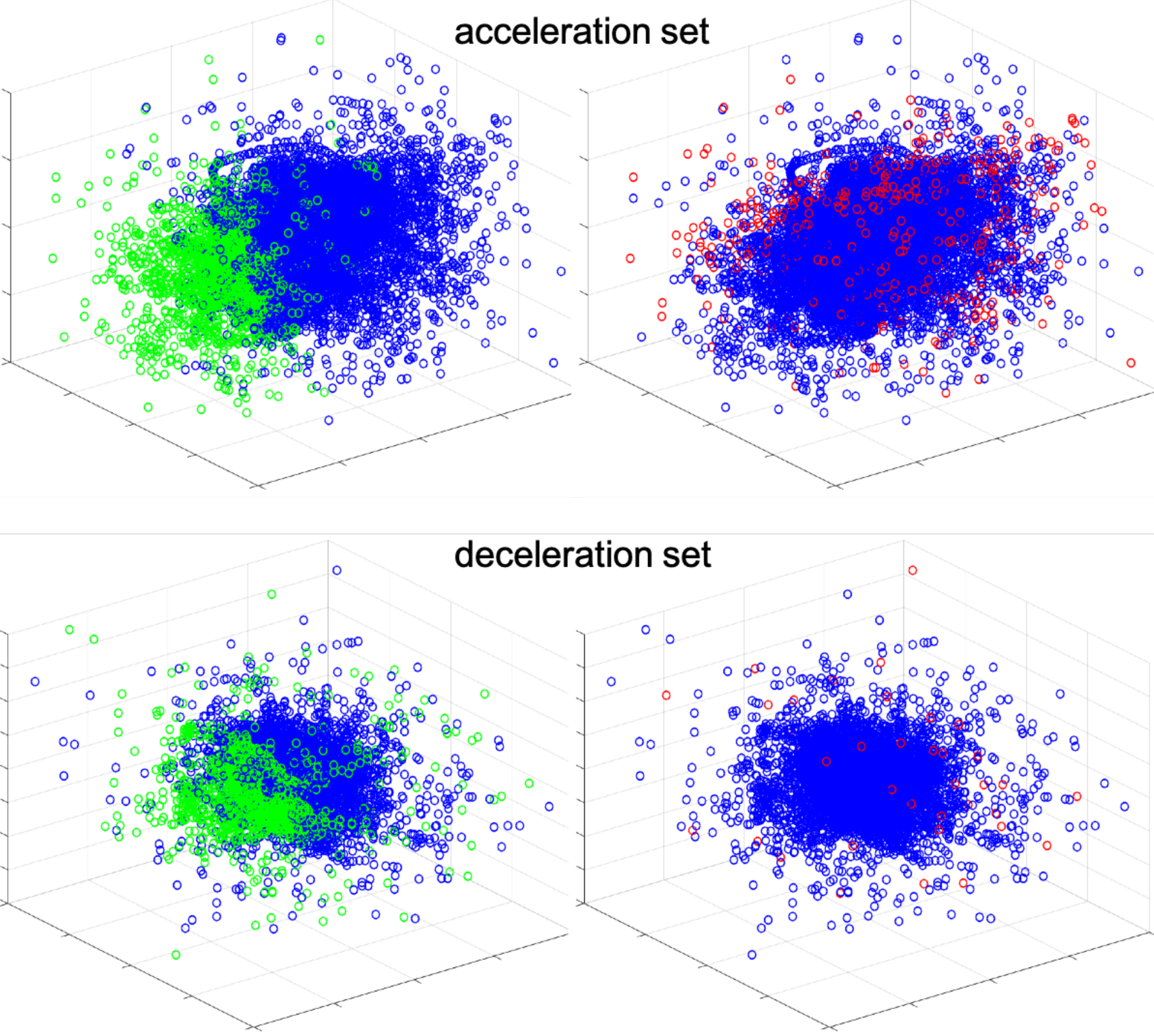}}
	\caption{Left subfigure: predicted genuine (blue) and trigger (green) samples. Right subfigure: the ground truth genuine (blue) and trigger samples (red). (a) Congestion attack in single-lane circuit, (b) insurance attack in single-lane circuit.}
	\label{defense_clustering}
\end{figure}
to see if the activations from the trigger samples and the genuine samples are distinguishable.  We present the results in Fig.~\ref{defense_clustering}. We find that trigger samples cannot be distinguished from the genuine samples using activation clustering. We also use the Silhouette score as suggested by authors in \cite{Activation_clustering} to determine whether the data is poisoned. We calculate the Silhouette score for both the poisoned data and the genuine data for two clusters. A significantly lower Silhouette score for the genuine data and significantly higher score for the poisoned data indicate that two clusters fit the poisoned data better. In our example, the Silhouette scores for the poisoned data and genuine data are comparable, e.g., 0.74 for the genuine and 0.75 for the poisoned, which further shows that activation clustering algorithm cannot detect the trigger samples generated by our method.

\section{Conclusion}\label{s:conclusion}
In this work, we propose attacks on DRL-based controllers for AVs by trojanning the machine learning models. Using specific combinations of sensor measurements as triggers, we were able to stimulate the maliciously trained neurons at the precise moment of attack. Since those malicious neurons do not interfere  with the normal functioning of the controllers, they remain undetected during benign operation. We analyze the stealth of the triggers using a measure of distance between genuine samples and the trigger samples in the pre-injection stage. We test the backdoor attacks in a single-lane circuit and a two-lane with an AV and observe in both cases that the backdoored model can successfully produce stop-and-go traffic congestion and crashes when triggered, and accomplish the benign control objective of removing the stop-and-go waves when not triggered. Contrary to the literature discussing backdoors in machine learning-based classification models, our triggers are not modular manipulations to the images  (like sun-glasses or post-its), which may be physically removed.  We, thus, tested our triggers using two latent-space detection techniques and demonstrated that neither was able to detect our triggers. Hence, state-of-the-art defenses are not suitable for the kinds of attacks we developed. We conclude that for AVs controlled by DRL-based controllers, there is a need for efficient backdoor detection and suppression.

\section*{ACKNOWLEDGMENT}
This work was jointly supported by the NYUAD Center for Interacting Urban Networks (CITIES), funded by Tamkeen under the NYUAD Research
Institute Award CG001 and by the Swiss Re Institute under the Quantum Cities™ initiative, and Center for CyberSecurity (CCS), funded by Tamkeen under the NYUAD Research Institute Award G1104.

%\section*{Resources}
%The simulation model and the trigger selection methodology will be open sourced as benchmarks for defense development.

\appendix
\gdef\thesection{Appendix \Alph{section}}

%\section*{References}
	
	%% References
	
	
\bibliographystyle{plainnat}
\bibliography{References_Saif}
	
	
	
	
	
	
	% \end{thebibliography}

\end{document}